\documentclass[preprint,10pt]{elsarticle}
\usepackage[margin=1in]{geometry}
\usepackage[utf8]{inputenc}
\pdfoutput=1 
\usepackage{changepage}

\usepackage{graphicx}
\usepackage{float}

\usepackage{amsmath}
\usepackage{amssymb}
\usepackage{amsthm}
\usepackage{mathtools}
\usepackage{bm}
\usepackage{stmaryrd}

\usepackage{booktabs}
\usepackage{array}
\usepackage{tabularx}
\usepackage{longtable}

\usepackage{algorithm}
\usepackage[noend]{algpseudocode}

\usepackage{tikz}
\usetikzlibrary{shapes,arrows,positioning,chains,fit,arrows.meta,calc}
\usepackage{subcaption}

\usepackage{pgfplots}
\usepackage{pgfplotstable}
\pgfplotsset{compat=1.5}

\usepackage{xcolor}
\usepackage{siunitx}
\usepackage{xspace}
\usepackage{comment}

\usepackage{hyperref}
\usepackage{cleveref}

\journal{}

\begin{document}
\makeatletter
\def\ps@pprintTitle{%
  \let\@oddhead\@empty
  \let\@evenhead\@empty
  \let\@oddfoot\@empty
  \let\@evenfoot\@oddfoot
}
\makeatother
\begin{frontmatter}
\title{
Multi-Agent Digital Twins for \\ Strategic Decision-Making using Active Inference
}

\author[mox]{Francesco Maria Mancinelli}
\author[dica]{Matteo Torzoni}
\author[cnr]{Domenico Maisto}
\author[cnr]{Francesco Donnarumma}
\author[dica]{\\ Alberto Corigliano}
\author[cnr]{Giovanni Pezzulo}
\author[mox]{Andrea Manzoni}

\address[mox]{MOX -- Department of Mathematics, Politecnico di Milano, Milan, Italy}
\address[dica]{Department of Civil and Environmental Engineering, Politecnico di Milano, Milan, Italy}
\address[cnr]{Institute of Cognitive Sciences and Technologies, National Research Council, Rome, Italy}

\begin{abstract}
Active Inference is an emerging framework providing a quantitative account of behavioral processes in neuroscience and a principled approach to decision-making under uncertainty. Its application to agency problems is natural, offering an autopoietic interpretation of action while addressing classical challenges such as the exploration--exploitation trade-off. Recently, Active Inference has been applied to digital twin scenarios for adaptive and predictive modeling of complex systems. In this work, we extend Active Inference to multi-agent digital twins in which agents interact within a shared environment while maintaining decentralized generative models. Our multi-agent framework features two innovations: (i) contextual inference to improve adaptability in dynamic environments, and (ii) the integration of streaming machine learning within agents' generative structures, enabling tunable goal-oriented behavior while preserving efficiency and scalability. The framework is illustrated through a Cournot competition example, providing a digital twin representation of a socio-economic system and highlighting its potential for coordinated decision-making in multi-agent contexts.
\end{abstract}


\end{frontmatter}

\section{Introduction}

Active Inference (AIF) is a neuroscience-inspired framework for decision-making under uncertainty. Originally developed for theoretical neurobiology and computational psychiatry~\cite{ActInfBook, psychosis}, AIF has demonstrated significant effectiveness in modeling adaptive behavior in biological systems~\cite{psychosis, CULLEN2018809, Pesci, uccellini, maisto2025flock}, as well as in informatics and engineering-related disciplines~\cite{IoT, Orchestrator}. It offers a unified Bayesian approach that integrates perception, action, and learning, enabling autonomous agents to operate effectively in partially observable environments. The probabilistic foundation of AIF makes it particularly well suited for predictive decision-making and adaptive control scenarios, where uncertainty and partial observability are often inherent challenges. On the other hand, digital twins (DTs) are nowadays widely adopted as virtual representations of physical systems that are continuously updated through real-time data to support monitoring, prediction, and control. They have been applied across domains such as manufacturing, automotive, aerospace, personalized medicine, and smart cities~\cite{DTmedicine, DTgeneral, SmartCity}. Building upon a recently proposed AIF-based DT framework~\cite{MT_AIF}, this work proposes a multi-agent extension for systems of interacting entities, where coordination emerges from decentralized inference.

\begin{figure}[t]
    \centering
    \includegraphics[width=0.8\linewidth]{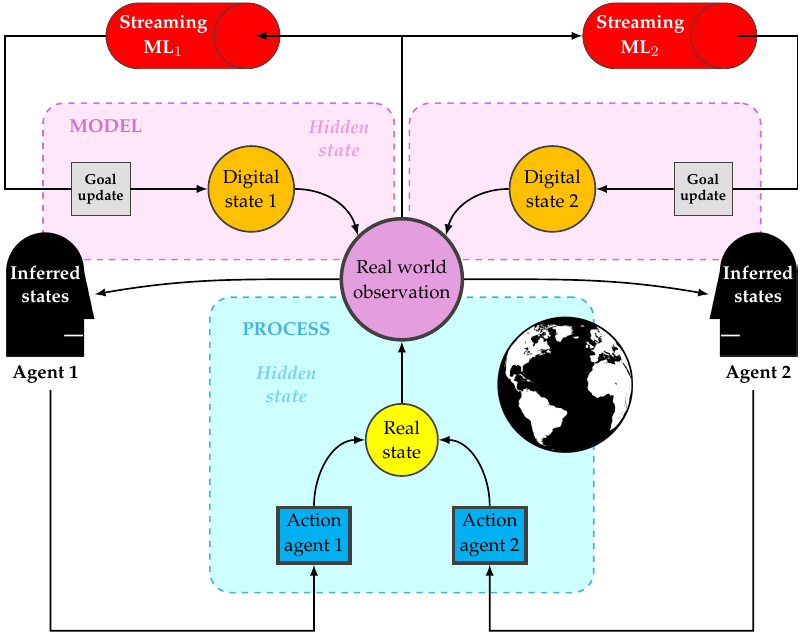}
    \caption{Generative process (GP) and generative model (GM) in the proposed multi-agent framework. The GP represents the true causal structure of the environment that produces observations, while the GM is the internal probabilistic model used by each agent to perform inference and guide actions. Arrows denote causal relations. Actions generated through inference affect the GP, closing the action-perception loop. In this work, GMs are supported by streaming machine learning tools.}
    \label{fig:GPvsGM}
    \vspace{-0.25cm}
\end{figure}

Inspired by self-organizing biological systems, as seen in morphogenesis~\cite{morphogenesis}, AIF enables decentralized coordination through probabilistic belief updating. Each agent operates via its generative model (GM), a probabilistic, parameterized representation of how hidden states of the environment generate observable outcomes. While the GM corresponds to the internal probabilistic model used by each agent to explain evidence and guide decision-making, the actual causal structure of the environment is referred to as the generative process (GP). Figure~\ref{fig:GPvsGM} illustrates this relationship in the proposed multi-agent setting: agents maintain their own GM and perform inference on hidden states while accounting for the presence of others via shared observations from a common environment.

From the perspective of DTs development, AIF offers several potential advantages. Unlike reinforcement learning approaches, which often rely on trial-and-error exploration and large datasets, AIF agents use compact GMs to infer hidden states and plan actions. The purpose of these models is not necessarily to accurately reconstruct the environment, but rather to provide a functional representation that allows agents to predict observations and respond adaptively to environmental dynamics. Moreover, decision-making in AIF naturally arises from balancing pragmatic objectives (achieving preferred outcomes) and epistemic drives (reducing uncertainty through information-seeking)~\cite{NoCostFunction}, enabling goal-directed exploration without externally imposed information-seeking strategies.

In many real-world applications, agents operate under non-stationary conditions, where both environmental dynamics and the desirability of outcomes may evolve over time. Within the streaming machine learning (SML) literature, this setting is typically addressed under the notion of concept drift, requiring models to adapt continuously under memory and computational constraints~\cite{gama2014survey, lu2018learning}. While most existing approaches focus on adapting predictive models, comparatively less attention has been devoted to the online adaptation of value or preference structures when outcome desirability is itself non-stationary or can not be fully captured by latent contextual variables~\cite{pezzulo2018hierarchical,Sajid_2022}. This limitation is particularly relevant in AIF, where prior preferences over outcomes play a central role in guiding policy selection through expected free energy minimization~\cite{friston2017active, parr2019generalised}. Although these prior preferences have been linked to reward signals in reinforcement learning~\cite{tschantz2020reinforcement}, they are typically assumed to be fixed or to follow predefined schedules. As a result, the problem of adapting preferences online remains largely unexplored.

This work proposes a multi-agent AIF framework for DTs in which agents interact within a shared environment through decentralized GMs with tunable goal-oriented behavior. While the AIF paradigm has already been widely applied within multi-agent frameworks~\cite{Pesci, uccellini, Orchestrator, morphogenesis, maisto2025flock, AIF_GT, IoT, Bottoni}, its use for the development of DTs has not yet been explored. The proposed framework therefore extends multi-agent AIF to DT settings by introducing two methodological novelties: ($i$) it provides DTs with mechanisms for detecting environmental changes and enabling context-sensitive adaptation; and ($ii$) it integrates SML techniques to support adaptive goal-directed behavior when outcome values evolve over time.

Specifically, we model preferences as dynamic quantities inferred online from interaction data via SML, allowing agents to adapt their goal structures in non-stationary environments. Importantly, updating preferences does not replace contextual inference; rather, it complements it by enabling agents to adapt not only their beliefs about hidden states and environmental dynamics, but also their preferences over outcomes. This results in a form of adaptive preference learning, i.e., online adaptation of the pragmatic component of expected free energy, that operates alongside standard epistemic updates in AIF and is particularly relevant in multi-agent and DT settings where objectives may evolve over time.

To illustrate the proposed approach, we consider an extended Cournot competition model, a classical game-theoretic framework in which firms compete by choosing production quantities in an oligopolistic market. In the proposed setting, each firm is represented by a decision-making agent responsible for production and warehouse management over time. This scenario serves as a testbed for studying the interaction of multiple AIF agents engaged in competitive decision-making and allows for comparison with traditional equilibrium-based results. While different from classical applications of the DT paradigm, the considered scenario can be fully interpreted as a DT representation of a socio-economic system -- in line with the definition\footnote{A digital twin is a set of virtual information constructs that mimics the structure, context, and behavior of a natural, engineered, or social system (or system-of-systems), is dynamically updated with data from its physical twin, has a predictive capability, and informs decisions that realize value.}. Each firm operates as a virtual counterpart of a socio-economic operator embedded in an evolving environment. Within this framework, the model also allows the investigation of collective behavioral phenomena such as imitation effects or bandwagon dynamics~\cite{1950Bandwagon}.

The manuscript is organized as follows. Section~\ref{sec:AIagents} introduces AIF agents, the corresponding generative modeling framework, and the integration of SML components. Section~\ref{sec:num_exp_1} presents the Cournot-based scenario and discusses the numerical experiments. Finally, Section~\ref{sec:Conclusions} summarizes the main conclusions.

\section{Active Inference Agents in POMDPs}
\label{sec:AIagents}

Active Inference provides a principled framework for modeling perception, action, and learning as processes of probabilistic inference. In this perspective, agents rely on a GM, an internal probabilistic representation that encodes beliefs about hidden states, their temporal evolution, and the observations they generate, in order to infer the latent causes of sensory data. This internal model differs from the GP, which represents the true external dynamics of the environment producing observations. The discrepancy between these two structures requires agents to continuously update their beliefs by minimizing variational free energy (VFE), a quantity that bounds the surprise associated with incoming observations.

When expressed within the partially observable Markov decision process (POMDP) formalism, the GM includes hidden states, observation likelihoods, transition probabilities, and prior preferences over outcomes. This representation enables the integration of perception and decision-making within a single probabilistic framework, where action selection emerges from the minimization of expected free energy (EFE) and naturally incorporates both goal-directed and information-seeking behaviors~\cite{knowingOnesPlace}. In the following, we introduce the main elements of this formulation and discuss how they support the development of adaptive agents. We then describe how AIF agents for DTs can exhibit context-sensitive behavior and, finally, how SML techniques can be integrated to modulate goal priors online.

\subsection{The POMDP Problem} 
\label{sec:POMDP_problem}

Partially observable Markov decision processes provide a formalism for modeling sequential decision-making problems under uncertainty, in settings where the full state of the environment is not directly accessible. Agents must act optimally based on uncertain and incomplete sensory data, leveraging a probabilistic model to infer latent states and select control policies (or actions) defining a long-term behavior.

Formally, a POMDP comprises a tuple $(\mathcal{D}, \mathcal{O}, \mathcal{U}, \mathcal{R}, \mathbf{A}, \mathbf{B}, \boldsymbol{\phi})$, where: $\mathcal{D}$ denotes a set of latent, digital states, designed to capture the essential features of the real-world hidden states, whose space is referred to as $\mathcal{S}$; $\mathcal{U}$ is a set of actions or control states; $\mathcal{O}$ is a set of observable outcomes; $\mathbf{B}$ is the transition model encoding the dynamics of state evolution from $d\in \mathcal{D}$ to $d'\in \mathcal{D}$ under action $u \in \mathcal{U}$; $\mathbf{A}$ is the observation model specifying the probabilistic mapping from hidden states $d$ to observations $o \in \mathcal{O}$; $\mathcal{R}(d, u)$ is the reward function that defines the agent's preferences; and $\boldsymbol{\phi}$ denotes the set of model parameters governing the probabilistic mappings (e.g., the parameters of the transition and observation models). We assume that digital states, observational data, and control actions are defined over discrete and finite spaces. This implies that each of these variables can only take values in a finite set of discrete levels. A natural representation for the associated probability distributions is given by categorical distributions. These assign a probability value between 0 and 1 to each discrete outcome, under the constraint that probabilities across all levels sum to one, as they represent a complete and mutually exclusive set of realizations.

Let $D_t$, $O_t$, $U_t$, and $R_t$ denote the random variables taking values in the corresponding sets $\mathcal{D}$, $\mathcal{O}$, $\mathcal{U}$, and $\mathcal{R}$ at time $t$. The joint probability distribution $p(O_t, D_t, U_t, R_t, \boldsymbol{\phi})$ therefore describes the probabilistic structure of the POMDP at each time step. In practice, these discrete probability distributions are represented through conditional probability tables (CPTs). Within the AIF formulation, these CPTs correspond to the matrices that parameterize the GM and will be described explicitly in Section~\ref{sec:AIF_GM}. In this sense, the POMDP specification provides the probabilistic framework from which the GM used by the agent is constructed.

\begin{figure}
    \centering
    \includegraphics[width=0.8\linewidth]{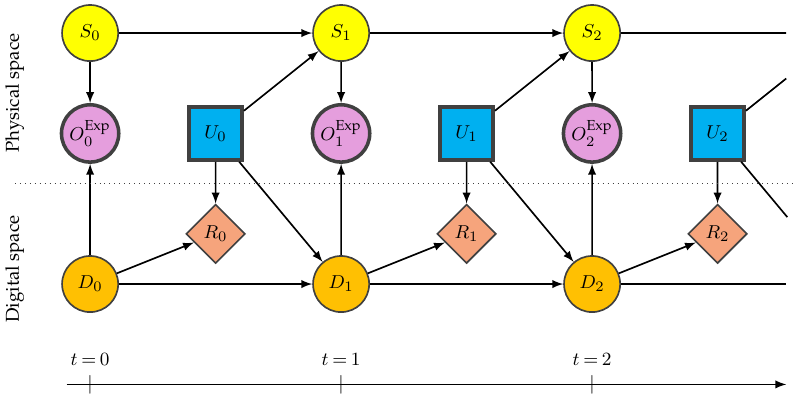}
    \caption{Dynamic Bayesian network encoding the relations among the real-world physical space and its digital representation. Circular nodes represent random variables, square nodes denote taken actions, diamond-shaped nodes symbolize the objective function.Bold nodes are observed quantities, thin nodes are latent. Directed edges encode conditional dependencies.}
    \label{fig:DBN_POMDP}
\end{figure}

A graphical representation of the POMDP structure is shown in Figure~\ref{fig:DBN_POMDP}, using a dynamic Bayesian network (DBN) to capture the temporal and probabilistic dependencies among the actual and inferred components of the system. In this representation, circular nodes indicate random variables, square nodes correspond to actions taken by the agent, and diamond-shaped nodes represent the objective function. These elements are indexed by discrete time steps $t \in \{0, \ldots, T\}$, where $t = 0$ denotes the beginning of the process and $t = T$ its operational horizon. 
Nodes with bold outlines indicate observed variables; those with thin outlines denote latent variables requiring inference. Edges between nodes express conditional dependencies, yielding a sparse structure that reflects the causal and informational relationships across time.

The physical-to-digital information flow from the environment to the agent is mediated by observations. The physical state $S_t$ produces an observed outcome $O_t^{\rm Exp}$ (the experienced observation), which constitutes the sensory data available to the agent. At the same time, the agent's GM specifies a probability distribution over possible observations $O_t$, conditioned on its beliefs about the digital state $D_t$. In this sense, the upward arrow from $D_t$ to $O_t^{\rm Exp}$ reflects the inferential process through which sensory evidence updates the agent's beliefs about the underlying digital state. The experienced observation $O_t^{\rm Exp}$ is compared with $O_t$, and the resulting mismatch drives the update of the posterior beliefs encoded in $D_t$. Since the physical state $S_t$ is only partially and indirectly observable, the digital state $D_t$ encodes the posterior beliefs over possible digital state configurations at time $t$, reflecting the evidence provided by the available observations. The observation space $\mathcal{O}$ may include sensor recordings, inspection results, or diagnostic information. 

\subsection{Active Inference Generative Model}
\label{sec:AIF_GM}

\begin{figure}
    \centering
    \includegraphics[width=0.9\linewidth]{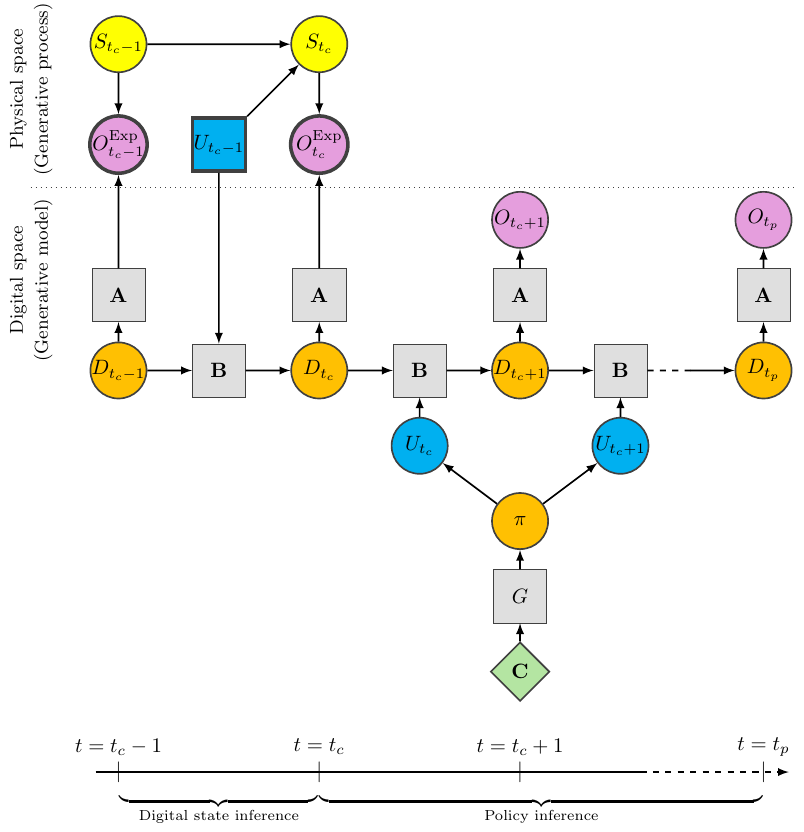}
    \caption{Dynamic bayesian network for the active inference generative model underlying the digital twin problem. Circular nodes denote random variables; blue squares actions; gray squares generative model operators; diamond node prior preferences. Bold nodes are observed, thin nodes are latent. Directed edges encode conditional dependencies. The upper path shows the physical space, the lower the digital inferred space. State inference occurs at current time $t_c$, policy inference propagates the digital state to a future time $t_p$.}
    \label{fig:DBN_AIF_Policy}
\end{figure}

Generative models of AIF agents are an internal probabilistic representation of the environment, useful to infer hidden causes of observations, predict outcomes, and select actions that fulfill goals. From a computational perspective, discrete GMs comprise four main components, expressed as categorical distributions and capturing different aspects of the environment~\cite{pymdp}: the observation likelihood $\mathbf{A}$, corresponding to $p(O_t \mid D_t)$, specifying how observations are probabilistically generated from hidden states; the transition likelihood $\mathbf{B}$, defined as $p(D_{t+1} \mid D_t, u_t)$, describing the evolution of hidden states over time as a function of previous states and control actions; the prior preferences $\mathbf{C}$, associated with $p(O_t)$, encoding the agent’s desired outcomes by assigning higher probability to preferred observations over time; and the initial state prior $\mathbf{D}$, given by $p(D_0)$, representing prior beliefs about the initial environmental state before any observation is incorporated. The bottom side of Figure~\ref{fig:DBN_AIF_Policy} provides a graphical representation of the DBN encoding the GM; additional details of the figure are discussed in Section~\ref{sec:epistemic}.

The AIF framework introduces two main modifications to the general POMDP formulation. First, the action variable $u$ is replaced by a policy $\pi = \lbrace u_{t_c},...,u_{t_d}\rbrace$. Policies are treated as latent variables to be inferred and, consistently with this interpretation, are represented as circular nodes in the diagram. The GM is typically presented as conditioned on a fixed policy $\pi$, which is how it is used for inference purposes. The posterior over policies represents the agent’s internal beliefs about its intended actions, while individual actions are interpreted as realizations sampled from the posterior over control states. The policy-to-control mapping $p(U_t | \pi)$ assigns at each time-step the appropriate control state based on the selected policy. The second modification concerns the omission of the reward variable $R$, as pragmatic goals are expressed through the prior distribution over future observations encoded in the unconditional CPT $\mathbf{C}$. These preferences may also vary over time to reflect evolving objectives, as discussed in Section~\ref{sec:SML}. Also graphically, the DBN reflecting the AIF perspective depicted in Figure~\ref{fig:DBN_AIF_Policy} shows substantial differences when compared with the general representation in Figure~\ref{fig:DBN_POMDP}. For time- (and space-) discrete POMDPs, probabilistic estimates of future states and observations over the prediction time steps $t = t_c, ..., t_p$ are computed as:
\begin{equation}
    p(O_{t_c:t_p}, D_{t_c:t_p}, \boldsymbol{\phi} | \pi) = p(\boldsymbol{\phi})p(D_{t_c};\boldsymbol{\phi})\prod_{t = t_c + 1}^{t_p} p(D_t|D_{t-1}, \pi; \boldsymbol{\phi}) \prod_{t = t_c}^{t_p} p(O_t|D_t; \boldsymbol{\phi}).
    \label{eq:AIF_POMDP}
\end{equation}

\subsection{State Inference via Free Energy Minimization}
\label{ssec:G}

Given an observation $O_{t_c}^{\text{Exp}} = o_{t_c}^{\text{Exp}}$, the agent must infer which digital state $D_{t_c}$ most likely explains the observation under its GM. This amounts to updating the agent’s belief distribution over hidden states in light of incoming data. Formally, this inference can be expressed through the posterior distribution over digital states:
\begin{equation}
p(D_{t_c} \mid O_{t_c}^{\text{Exp}} = o_{t_c}^{\text{Exp}})
=
\frac{p(o_{t_c}^{\text{Exp}} \mid D_{t_c})\, p(D_{t_c})}
{\sum_{d \in \mathcal{D}} p(o_{t_c}^{\text{Exp}} \mid d)\, p(d)},
\label{eq:State_Bayes}
\end{equation}
where the joint distribution $p(o_{t_c}^{\text{Exp}}, D_{t_c})$ factorizes into a likelihood term $p(o_{t_c}^{\text{Exp}} \mid D_{t_c})$ and a prior $p(D_{t_c})$. The denominator $p(o_{t_c}^{\text{Exp}})$ corresponds to the marginal likelihood, or model evidence, representing the observation probability under the GM after marginalizing over digital states.

In practice, computing the posterior distribution~\eqref{eq:State_Bayes} is often intractable because evaluating the marginal likelihood requires summing over all possible hidden state configurations. To address this challenge, AIF adopts variational inference~\cite{murphy2023probabilistic}, which approximates the true posterior $p(D_{t_c} \mid O_{0:t_c}^{\text{Exp}} = o_{0:t_c}^{\text{Exp}})$ with a tractable variational distribution $Q(D_{t_c}; \boldsymbol{\theta}) : \mathcal{D} \mapsto [0,1]$, parametrized by $\boldsymbol{\theta}$. This leads to the following optimization problem:
\begin{equation}
\boldsymbol{\theta}^* = \arg \min_{\boldsymbol{\theta}} D_{\text{KL}} \left[ Q(D_{t_c}; \boldsymbol{\theta}) \, \| \, p(D_{t_c} \mid o_{0:t_c}^{\text{Exp}}) \right],
\label{eq:theta_star}
\end{equation}
where $D_{\text{KL}}[Q(X) \| P(X \mid Y)] = \mathbb{E}_Q \left[\ln Q(X) - \ln P(X \mid Y)\right]$ denotes the Kullback–Leibler (KL) divergence between the approximate posterior $Q(X)$ and the true posterior $P(X \mid Y)$, for two generic random variables $X$ and $Y$. Here, $\mathbb{E}_Q$ denotes the expectation with respect to the variational posterior. 
Directly optimizing Eq.~\eqref{eq:theta_star} is however impractical because the true posterior distribution is unknown. To provide a tractable objective function for approximate Bayesian inference, we introduce the VFE as follows:
\begin{equation}
\mathcal{F}_{t_c}(\boldsymbol{\theta})
=
\mathbb{E}_Q \left[\ln Q(D_{t_c}; \boldsymbol{\theta}) - \ln p(o_{t_c}^{\text{Exp}}, D_{t_c}) \right].
\label{eq:VFE_F}
\end{equation}
As shown in \cite{MT_AIF}, minimizing VFE is equivalent to minimizing the KL divergence in Eq.~\eqref{eq:theta_star}. The resulting variational distribution $Q(D_{t_c}; \boldsymbol{\theta})$ assigns high probability to digital states that both (i) are consistent with the agent’s prior beliefs about system dynamics and (ii) provide a plausible explanation for the observed data. At convergence, if the variational posterior matches the true posterior, the KL divergence vanishes and the VFE reduces to the negative log marginal likelihood, also known as Bayesian surprise.

In sequential settings, as that defined by the GM in Eq.~\eqref{eq:AIF_POMDP}, inference must be performed each time new observations become available. The agent therefore updates its belief over the current digital state $D_{t_c}$ by combining the latest observation with prior beliefs propagated from the previous time step. Under this formulation, instantaneous inference involves approximating the following posterior distribution:
\begin{equation}p(D_{t_c} \mid O_{t_c} = o_{t_c}^{\text{Exp}}, D_{t_c-1}, U_{t_c-1} = u_{t_c-1}).
\end{equation}
According to \cite{MT_AIF}, by using the VFE objective~\eqref{eq:VFE_F}, the inference problem can be written as:
\begin{equation}
\boldsymbol{\theta}^* 
= \arg \min_{\boldsymbol{\theta}} \, \mathbb{E}_Q \left[\ln Q(D_{t_c}; \boldsymbol{\theta}) - \ln \left(p(o_{t_c}^{\text{Exp}} \mid D_{t_c}; \boldsymbol{\phi}) p(D_{t_c} \mid D_{t_c-1}, u_{t_c-1}; \boldsymbol{\phi})\right)\right].
\label{eq:theta_explicit}
\end{equation}

The optimization problem in Eq.~\eqref{eq:theta_explicit} can be solved using fixed-point iteration methods \cite{GraphModAndVarInf}. Under the assumption of temporal factorization, variational posteriors at different time steps are conditionally independent. As a result, the free energy across a trajectory decomposes into a sum of single-time-step contributions, enabling independent inference updates at each time step. Moreover, the variational posterior $Q(D_{t_c}; \boldsymbol{\theta})$ at a given time step can be further factorized across $F$ independent hidden state factors $D = \{D^1,\ldots, D^F\}$ according to the mean-field approximation:
\begin{equation}
Q(D_{t_c}; \boldsymbol{\theta}) = \prod_{f=1}^{F} Q(D_{t_c}^f; \boldsymbol{\theta}),
\label{eq:QoI_fact}
\end{equation}
where $Q(D_{t_c}^f; \boldsymbol{\theta})$ denotes the posterior over the $f$-th hidden state factor, $f = 1, \ldots, F$. Each of these factors may represent distinct aspects of the GP, potentially varying in dimensionality, transition dynamics, and association with specific observation modalities. Similarly, observations can be structured into $M$ distinct modalities $O = \{O^1, \ldots, O^M\}$, where each $O^m, m = 1, \ldots, M$, corresponds to a separate sensory channel used by the agent at each time step.

In this multi-modal, multi-factor setup, the observation likelihood array $\mathbf{A}$ becomes a collection of $M$ sub-arrays $\mathbf{A} = \{\mathbf{A}^1, \ldots, \mathbf{A}^M\}$, with each $\mathbf{A}^m, m = 1, \ldots, M$, representing the observation model for the $m$-th modality. Each sub-array encodes the likelihood $p(O^m \mid D^1, \ldots, D^F; \boldsymbol{\phi})$, capturing the dependency of that observation modality on the hidden state factors. Similarly, the transition model $\mathbf{B}$ is represented as a collection of $F$ sub-arrays $\mathbf{B} = \{\mathbf{B}^1, \ldots, \mathbf{B}^F\}$. Each $\mathbf{B}^f, f = 1, \ldots, F$, encodes the dynamics $p(D_t^f \mid D_{t-1}^f, u_{t-1}^f; \boldsymbol{\phi})$, conditioned on the previous state and action for that factor. Control states are factorized analogously to hidden states, such that $U = \{U^1, \ldots, U^F\}$. Each control factor $U^f$ governs the transitions of the corresponding digital state factor $D^f$, with dimensionality matching the number of possible control actions for that system aspect. 

\subsection{Epistemic Behavior Drives Information} \label{sec:epistemic}

Given the updated variational posterior over the digital state $Q(D_{t_c}; \boldsymbol{\theta})$, AIF enables policy inference by evaluating the quality of each admissible policy $\pi$ comprising sequences of future actions over a prediction horizon $t = t_c, \ldots, t_p$. Central to this process is the EFE, a quasi-utility function that can be interpreted as the expected value of the VFE (see Eq.~\eqref{eq:VFE_F}), under predicted future trajectories generated by a candidate policy. It provides a tractable objective for evaluating policies that unfold over time, by scoring their consequences before observations are actually realized. While resembling the VFE in form, the EFE evaluates future trajectories in terms of both pragmatic and epistemic behaviors, namely the pursuit of preferred outcomes and the resolution of uncertainty. Consequently, it involves expectations over future hidden states and future observations under a given policy, as illustrated in Figure~\ref{fig:DBN_AIF_Policy}.

The EFE associated with a generic policy $\pi$ is defined as:
\begin{equation}
    G^{\pi} = \mathbb{E}_{Q(O_{t_c:t_p}, D_{t_c:t_p}|\pi)}\left[\log Q(D_{t_c:t_p}|\pi)- \log \tilde{p}(O_{t_c:t_p}, D_{t_c:t_p}|\pi)\right],
    \label{eq:EFE_base}
\end{equation}
where we omit for simplicity the explicit dependence of the variational posterior on the parameters $\boldsymbol{\theta}$ and the GM on hyperparameters $\boldsymbol{\phi}$. The biased GM $\tilde{p}(O_t, D_t|\pi) = p(D_t|O_t, \pi)\tilde{p}(O_t)$ integrates the prior preferences over observations encoded in $\mathbf{C}$, thus shaping behavior toward desirable outcomes. Policy inference is then cast as the following optimization problem:
\begin{equation}
    \pi^* = \arg\min_{\pi} G(\pi),
\end{equation}
where the lower the EFE $G(\pi)$, the higher the posterior probability assigned to the policy $\pi$. 

In practice, policy selection consists of two steps. First, for each candidate policy $\pi$, a quality score is assigned based on the negative EFE. Second, a policy is sampled or selected according to its EFE-minimization capability. To gain deeper interpretability into how policies are ranked,  Eq.~\eqref{eq:EFE_base} can be expanded at a single time step $t \in \{t_c, \ldots, t_p\}$, as follows:
\begin{equation}
\begin{aligned}
    G_t^\pi &= \mathbb{E}_{Q(O_t, D_t \mid \pi)} 
    \left[\ln Q(D_t \mid \pi) - \ln \tilde{p}(O_t, D_t \mid \pi)\right] \\
    &= - \underbrace{\mathbb{E}_{Q(O_t \mid \pi)} 
    \left[\text{D}_{\text{KL}}\left[Q(D_t \mid O_t, \pi) \,\|\, Q(D_t \mid \pi)\right]\right]}_{
    \text{Epistemic value (information gain)}} - \underbrace{\mathbb{E}_{Q(O_t \mid \pi)}
    \left[\ln \tilde{p}(O_t)\right]}_{
    \text{Pragmatic value (utility)}} \\
    &\quad + \underbrace{\mathbb{E}_{Q(O_t \mid \pi)} 
    \left[\text{D}_{\text{KL}}\left[Q(D_t \mid O_t, \pi) \,\|\, 
    p(D_t \mid O_t, \pi)\right]\right]}_{
    \text{Expected variational approximation error } (\geq 0)}.
\end{aligned}
\label{eq:EFE_unpacked}
\end{equation}
This EFE decomposition makes explicit the balance between exploration and exploitation. The \textit{epistemic value} term captures expected information gain. It encourages the selection of policies that are expected to reduce uncertainty about hidden states. Intuitively, this corresponds to maximizing the divergence between prior and posterior beliefs under policy $\pi$, and thus promotes the active sampling of informative outcomes. Although counterintuitive, this mechanism penalizes ``many-to-one'' mappings from observations $O_{t_c:t_p}$ to hidden states $D_{t_c:t_p}$, which hinder precise inference. The \textit{pragmatic value} term reflects expected utility by incorporating prior preferences over observations $\tilde{p}(O_t)$, effectively steering the agent toward preferred outcomes. 

In environments characterized by high uncertainty, epistemic actions are initially favored to reduce ambiguity, followed by pragmatic actions once confidence has improved. This mirrors a core limitation of many reinforcement learning algorithms, which often struggle to explain or incorporate epistemic exploration systematically. By contrast, AIF not only explains \textit{why} exploration should occur, but also offers a principled way to design agents that naturally balance epistemic and pragmatic drives. Epistemic actions can be explicitly engineered in the GM and pursued when valuable, even at a cost.

\subsection{Contextual Inference for Behavior Adaptation}
\label{sec:ctx_inf}

AIF agents operate on latent states inferred from observations, which allows them to exhibit context-sensitive behavior. When environmental conditions or task demands change, agents update their beliefs about the current context and adjust their policy selection accordingly. This mechanism enables flexible and robust behavior, as agents continuously infer the most likely causes of their observations and act accordingly. As a result, different behavioral modes can naturally emerge from a single model, reflecting different beliefs about hidden states and policies.

Exploiting the GM factored structure, it is possible to define a procedure that guides the agent to recognize the context in which it is operating. Practically, this is achieved by introducing a dedicated hidden state factor \(D^{\text{ctx}}\), responsible for encoding contextual information. A corresponding observation modality \(m_{\text{ctx}}\) provides sensory evidence about this factor, with an associated likelihood array $\mathbf{A}^{m_\text{ctx}} = p(O^{\text{ctx}} \mid D^{\text{ctx}})$. During model inversion, observations from this modality update beliefs over the contextual hidden states according to
\begin{equation}
q(D^{\text{ctx}}_t) \propto p(O^{\text{ctx}}_t \mid D^{\text{ctx}}_t)\, p(D^{\text{ctx}}_t).
\end{equation}

Subsequent to contextual inference, the remaining $m \neq m_\text{ctx}$ observation models can be conditioned on the contextual factor, such that they explicitly depends on the inferred context. Equivalently, each likelihood can be interpreted as a context-indexed family of observation models, as follows:
\begin{equation}
\mathbf{A}^m_{d_{\text{ctx}}}
=
p(O^m \mid D^1,\dots,D^{\text{ctx}} = d_{\text{ctx}},\dots,D^F),
\end{equation}
where inference over \(D^{\text{ctx}}\) selects the observation model most consistent with the current context.

\subsection{Streaming Machine Learning for Predictive Goal-Oriented Behavior}
\label{sec:SML}

In the GM of an AIF agent, the $\mathbf{C}$ array encodes prior preferences over observations. These preferences are typically fixed in time or assumed to follow a known temporal evolution~\cite{ActInfBook,pymdp}. In this work, we relax this assumption by allowing prior preferences to evolve dynamically as new information becomes available.

Denoting the quantities that govern inference over hidden states at time step $t$ as \mbox{$z_t = g(O_t, q(D_t), U_{t-1})$}, an SML component is introduced to map this information into updated goal priors, as follows:
\begin{equation}
\mathbf{C}(t) = f_{\text{SML}}(z_{1:t}),
\label{eq:C_hat_estimate}
\end{equation}
where $f_{\text{SML}}$ denotes a model trained incrementally on the stream of data generated by the agent's interaction with the environment (see also Figure~\ref{fig:GPvsGM}). Streaming machine learning methods are designed for scenarios in which data arrive sequentially and the underlying relationships may evolve over time. Unlike batch learning approaches, SML models update their internal parameters incrementally as new data become available. Denoting by $\boldsymbol{\xi}_t$ the parameters of the streaming model, the parameter update is written as:
\begin{equation}
\boldsymbol{\xi}_{t+1} = \Psi(\boldsymbol{\xi}_t, z_t),
\label{eq:xi_update}
\end{equation}
where $\Psi(\cdot)$ represents the online learning rule. The exact form of $\Psi$ depends on the specific streaming model. The updated model then produces revised estimates of quantities that influence the agent's preferences, as encoded in $\mathbf{C}(t)$.

The SML component acts as an adaptive interface between the data stream generated by the GP and the prior preferences encoded in $\mathbf{C}$. The GM itself remains unchanged, while preferences over outcomes are continuously refined as new information becomes available. The reason behind this modeling choice is that certain environmental variables may not be explicitly represented as hidden states within the GM, despite influencing outcomes associated with goal-priors. Explicitly modeling many additional hidden states factors would increase both the design burden and the computational complexity of the model. By delegating the inference of such influences to the external streaming learner, the agent can adapt its prior preferences while preserving a relatively compact GM. Note that although we focus here on updating the preference vector $\mathbf{C}$, the same principle can be extended to the likelihood array $\mathbf{A}$ or the transition dynamics $\mathbf{B}$. 

\section{Numerical Experiment: The Cournot Competition}
\label{sec:num_exp_1}
In this section, we assess the proposed methodology on a Cournot competition framework. The Cournot competition models a market with $n$ firms producing an identical good. Each firm chooses the production quantity $q_i$ while facing a marginal production cost $c_i$ and no fixed costs. The market price $P$ is determined by the pricing function:
\begin{equation}
P = f(q_1,\dots,q_n) = a - b\left(\sum_{i=1}^n q_i\right),
\end{equation}
which depends solely on the total quantity supplied. Here, $a$ denotes the maximum price customers are willing to pay, and $b$ is the price sensitivity coefficient reflecting how price decreases with increasing supply.

In the classical formulation, the game is solved by computing the Nash equilibrium via best response (BR) dynamics, where each firm maximizes its own revenue. When the pricing function is known (i.e., $a$ and $b$ are given), the problem admits a closed-form solution:
\begin{equation}
BR_i(q_i) = q_i(q_{\smallsetminus i}) = \frac{a - b\sum_{j \smallsetminus i} q_j - c_i}{2b}, \quad \forall i \in \{1,\dots,n\},
\label{BR_dynamics}
\end{equation}
with the full derivation available in~\cite{fudenberg1991game}. The solution in Eq.~\eqref{BR_dynamics} applies only to the static, one-shot version of the model. Here, $\smallsetminus i$ denotes the set of all firms other than $i$, while $\sum_{j \smallsetminus i} q_j$ the aggregate production excluding firm $i$.

Inspired by~\cite{landscapeDynamics}, which introduces a dynamic extension of the framework, we investigate how an AIF agent can operate in a multi-step, multi-agent setting. Each firm retains the same objective as in the classical case, maximizing revenue through its production decision, but production must now be modulated at each time step, denoted as $q_i(t)$. At every step, customers access the market and purchase goods, thereby determining the market price. 

A key feature of our dynamic formulation is that unsold items are stored in a warehouse. However, the agent controls only production and does not directly manage the warehouse. Consequently, the level of stored items must be inferred by the agent, including for future time steps during policy evaluation (i.e., production planning). To support this inference process, the warehouse emits a signal at each step reporting the current stock level. We assume that the warehouse signal is noisy, providing an opportunity to model explicit epistemic behavior: agents may request a detailed, yet costly, analysis of the warehouse state. While large inventories increase operational costs due to additional workload, maintaining stock allows firms to respond quickly to sudden increases in demand. The observation related to warehouse occupancy is treated as a context-specific observation, allowing the agent to distinguish between two situations: acceptable production levels or the need to reduce production.

We also introduce the use of Streaming Random Patches (SRP)~\cite{SRP} as an SML model to dynamically infer the price of products at the next-step price $P_{t+1}$. This approach avoids explicitly inferring competitors’ production or stock levels, which would substantially increase the complexity of the agents’ GM. From an AIF perspective, this setting naturally emphasizes the exploration–exploitation trade-off. Each agent assumes only the minimal information required to compute the BR initially, without additional assumptions about the environment. Exploring the state space becomes essential for achieving economic gain.

Section~\ref{sec:AIF_framework} presents a well-posed parameterization of the generative model underpinning the AIF agents (Sections~\ref{subsec:GM_A}-\ref{subsec:GM_D}), followed by the introduction of the generative process (Section~\ref{subsec:GP_Cour}) and the resulting AIF loop (Section~\ref{subsec:AIF_loop}). Section~\ref{sec:Simulations} reports the simulation results for both the standard duopoly dynamics and a three-firm extension, and investigates the effect of introducing an ill-posed competitor into the system. Finally, Section~\ref{subsec:sim_conclusions} discusses the simulation outcomes.

\subsection{Active Inference Framework}
\label{sec:AIF_framework}

Observational data $O_t$ consists of four modalities: the number of items sold (ranging from 0 to 10); the previous production decision; a stock occupancy signal with four levels (0 ($0\%\text{--}30\%$), 1 ($31\%\text{--}50\%$), 2 ($51\%\text{--}80\%$), and 3 ($>81\%$)); and a binary indicator of whether a warehouse analysis was performed. The control variable $U_t$ comprises two factors: the production quantity (ranging from 0 to 6 items); and the binary decision indicating whether to perform a warehouse analysis. In the reported results, this epistemic action is denoted as \textit{DN} (\textit{Do Nothing}) when the warehouse analysis is not performed, and as \textit{Analysis} when the epistemic action is executed. The digital state $D_t$ comprises three hidden state factors: the inferred warehouse inventory; the production context (acceptable vs.\ to be reduced), and an epistemic state that indicates whether a warehouse analysis is required.

It is important to emphasize that the digital state, observation, and action spaces reflect subjective design choices. The same problem can be formulated from alternative yet plausible perspectives, potentially leading to equivalent results. Indeed, the variables of the GM are not required to mirror real-world variables in their full complexity. Rather, they should provide minimal internal representations that enable the agent to interpret observations and interact effectively with the GP. In this work, the GM design is guided by the objective of recovering the theoretical outcomes of the classical Cournot competition.

\subsubsection{Likelihood Matrices [A]}
\label{subsec:GM_A}

\begin{figure}[t]
    \centering
    \begin{minipage}[b]{0.45\textwidth}
        \centering
        \includegraphics[width=\textwidth]{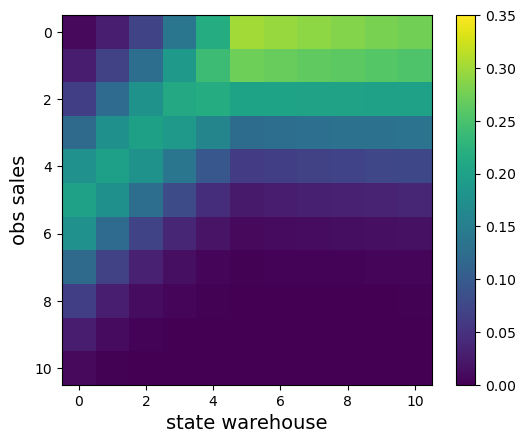}
        
        (a)
    \end{minipage}
    \hfill
    \begin{minipage}[b]{0.45\textwidth}
        \centering
        \includegraphics[width=\textwidth]{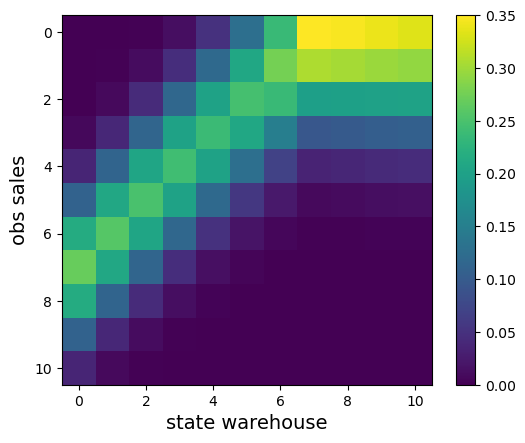}
        
        (b)
    \end{minipage}
    \caption{Sales likelihood matrices across production contexts: (a) acceptable production context; (b) reduced production context.}
    \label{fig:sales_likelihood}
\end{figure}

\paragraph{Sales likelihood:}
The sales likelihood array is constructed following the idea that the agent expects the warehouse to be empty when sales match the ``correct'' quantity, i.e., the sales amount that, according to prior analysis, corresponds to the BR. If observed sales fall short of this target, the probability that the warehouse is inferred to be increasingly full rises. Figure~\ref{fig:sales_likelihood} illustrates the two likelihood matrices associated with the ``acceptable production'' and ``reduce production'' contexts. The difference between these contexts is the number of sales needed to infer an empty warehouse; in the ``reduce production'' context, more sales are required. Consequently, for the same number of sales, this context leads to more conservative behavior, as the agent infers a higher number of unsold products in the warehouse.

\begin{figure}
    \centering
    \begin{minipage}[b]{0.45\textwidth}
        \centering
        \includegraphics[width=\textwidth]{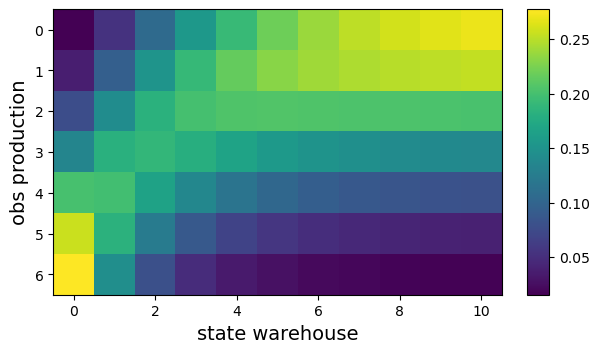}

        (a)
    \end{minipage}
    \hfill
    \begin{minipage}[b]{0.45\textwidth}
        \centering
        \includegraphics[width=\textwidth]{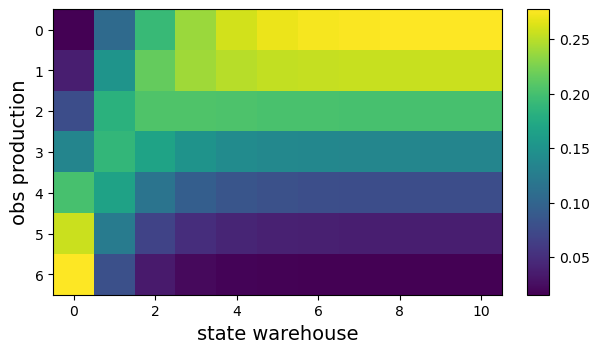}

        (b)
    \end{minipage}
    \caption{Production likelihood matrices across production contexts: (a) acceptable production context; (b) reduced production context.}
    \label{fig:production_likelihood}
    \vspace{-0.2cm}
\end{figure}

\paragraph{Production likelihood:}
The production likelihood is based on the assumption that agents tend to produce more when they infer the warehouse is empty. Figure~\ref{fig:production_likelihood} shows the two likelihood matrices associated with the two contextual assumptions. Under the ``reduce production'' context, the likelihood is adjusted so that low production levels lead the agent to estimate a larger number of stocked products, encouraging continued conservative behavior.

\paragraph{Likelihood for the stock level signal:}

Unlike the other likelihood arrays, the likelihood for the stock level signal requires combining two conditional likelihoods, each associated with a different hidden state: a context-specific slice ${p(\text{warehouse signal} \mid \text{warehouse})}$, which informs the agent’s estimation of the number of stocked products, and a context-dependent slice ${p(\text{warehouse signal} \mid \text{context})}$, which provides information useful for inferring the current context. See Appendix~\ref{app:signal_likelihood_derivation} for the full derivation. 

\begin{figure}
    \centering
    \begin{minipage}[b]{0.45\textwidth}
        \centering
        \includegraphics[width=\textwidth]{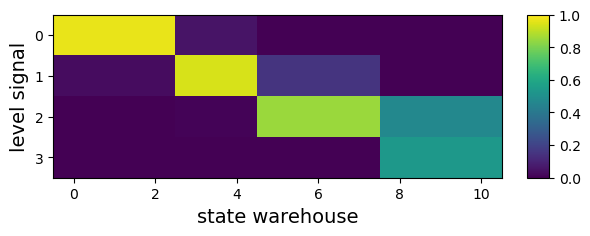}

        (a)
    \end{minipage}
    \hfill
    \begin{minipage}[b]{0.45\textwidth}
        \centering
        \includegraphics[width=\textwidth]{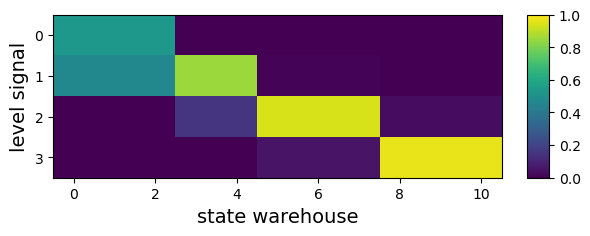}

        (b)
    \end{minipage}
    \caption{Context-sensitive likelihood ${p(\text{warehouse signal} \mid \text{warehouse})}$ under noisy conditions: (a) acceptable production context; (b) reduce production context.}
    \label{fig:signal_likelihood}
\end{figure}

Figure~\ref{fig:signal_likelihood} shows the noisy likelihoods ${p(\text{warehouse signal} \mid \text{warehouse})}$ under different contexts. The key intuition is that lower signal values are more likely when the warehouse contains fewer products, while higher values become more likely as the stock level increases. At the same time, the different signal patterns across contexts allow the agent to use this observation to update its beliefs about whether the system is in the \textit{acceptable production} or \textit{reduce production} context. Although the signal is inherently noisy, the agent can perform a costly epistemic (information-seeking) action to reduce uncertainty and obtain a more accurate signal.

\paragraph{Analysis likelihood:}

The likelihood associated with the epistemic state encodes whether a warehouse analysis has been performed. It reflects perfect introspective perception: when the analysis action is executed, the agent knows that its inference relies on a reliable, noiseless warehouse signal; otherwise, it relies on a corrupted observation. This likelihood does not represent uncertainty in the external environment but ensures internal consistency within the generative model. Specifically, it aligns the agent’s belief about the quality of sensory information with the selected epistemic action. This structure allows the agent to plan under different levels of observational uncertainty, balancing the cost of analysis against the benefit of obtaining a noiseless signal.

\subsubsection{Transition Matrices [B]}
\label{subsec:GM_B}

\begin{figure}
    \centering
    \vspace{-0.2cm}
    \includegraphics[width=1\textwidth]{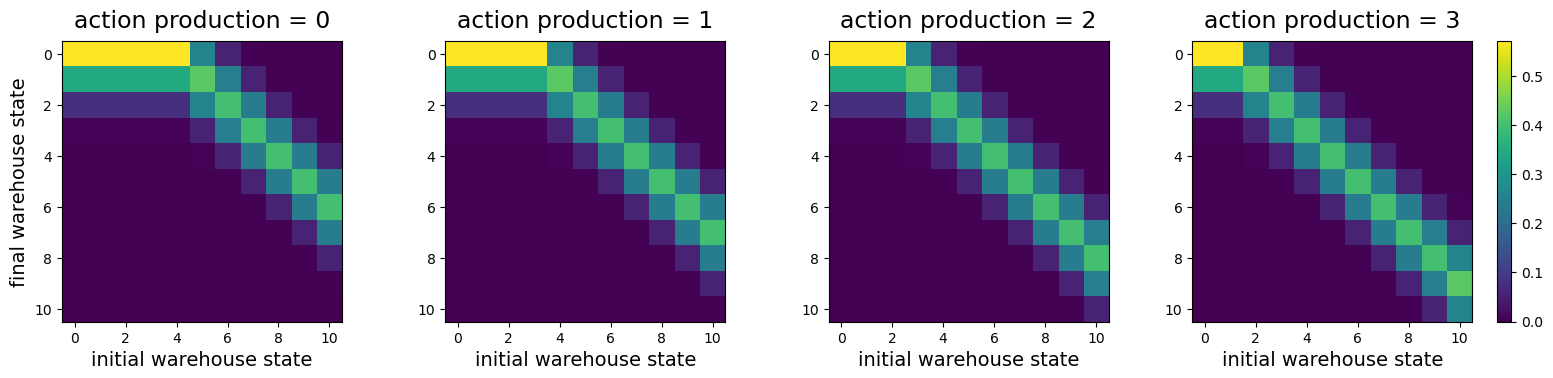}
    \caption{Example of warehouse transition assuming a one-step best response of 5 units.}
    \label{fig:warehouse_transition}
    \vspace{-0.15cm}
\end{figure}

\paragraph{Warehouse transition dynamics:}
The warehouse transition models the probabilistic one-step evolution of stock level given the agent’s production decision. It is calibrated using the firm's estimated BR, assuming that a sufficient number of customers will absorb the produced quantity. However, the BR is not independent of demand, as the market may not absorb the full production. This need for adaptivity is addressed using an SRP model: at each time step, the maximum market price (parameter $a$) is estimated; if the relative difference (in percentage terms) between the previous and updated estimates exceeds a threshold, the BR is recalculated. In practice, this updates the transition matrix by re-evaluating the BR using Eq.~\eqref{BR_dynamics}. Figure~\ref{fig:warehouse_transition} shows an example of the warehouse transition matrix for a BR corresponding to a production (and expected sales) of 5 units.

\paragraph{Context transition dynamics:}

The context transition encodes the agent's belief about the volatility of the production context. Since no action directly triggers a shift between contexts (i.e., from ``acceptable production'' to``reduce production'' or vice versa), this transition is considered passive. Its purpose is to capture the temporal evolution of the environment, independent of the agent's control. This transition probability is set uniformly to $\frac{1}{2}$, representing a neutral belief: at each time step, the agent assigns equal probability to remaining in the current context or switching to the other. While simple, this choice allows agents adaptivity without imposing strong prior assumptions. More complex transition models could be adopted to reflect specific environmental patterns or learned dynamics, but this is left for future work.

\paragraph{Analysis transition dynamics:}

The market analysis transition represents an agent's introspective state, depending on whether an epistemic (information-seeking) action has been performed. In line with the analysis likelihood, this transition is deterministic: if the agent performs a warehouse analysis, it enters an ``epistemic state'' with improved perceptual precision (i.e., noiseless warehouse signals). Otherwise, the agent remains in a non-epistemic state, relying on noisy observations.

\subsubsection{Costs and Prior Preferences [C]}
\label{subsec:GM_C}
\paragraph{Sales preferences:}

Prior preferences over sales encode the firm's desirability of expected monetary gains from selling goods. In the Cournot framework, prices depend on the total quantity supplied by the $n$ firms. Accordingly, the preference vector over sales at time $t$ is defined as
\begin{equation}
    \mathbf{C_{\text{sales}}}(t) = \hat{P}(q_1(t) + w_1(t), \dots, q_n(t) + w_n(t)) \cdot \overrightarrow{\text{sales}}_i,
\end{equation}
where $q_i(t)$ is the quantity produced by the $i$-th firm, with $i=1,\dots,n$, at time $t$, and $w_i(t)$ is its current warehouse stock, such that $q_i(t) + w_i(t)$ represents the total quantity available from the firm; $\overrightarrow{\text{sales}}_i$ denotes the possible sales levels; and $\hat{P}(\cdot)$ is the estimated market price provided by the SRP model (see Appendix~\ref{sec:SRP}), which returns a scalar estimate of the unit price given the quantities supplied by all firms. 

Since agents aim to match production to customer demand, we write $q_i(t) + w_i(t) \approx \text{sales}_i(t)$ to reflect the assumption that each firm attempts to meet demand exactly using both new production and existing inventory. In practice, the actual $\text{sales}_i(t)$ are directly observable by each firm, whereas the opponent's production $q_{\smallsetminus i}(t)$ and inventory $w_{\smallsetminus i}(t)$ are not. Accurately estimating them would require maintaining hidden state variables for the opponent's behavior and modeling their temporal dynamic, significantly increasing the complexity of the GM. To balance complexity and tractability, this work adopts a simplifying assumption: the firm estimates future prices using its own production levels and the opponent's sales from the previous time step. This is modeled as
\begin{equation}
    \hat{P}(q_i(t)) = f_{\text{SRP}}(q_i(t), \text{sales}_{\smallsetminus i}(t-1)).
    \label{eq:P_hat_estimate}
\end{equation}
At each time step, the firm computes $\hat{P}$ for all possible production quantities it may choose. These estimated prices are then multiplied by the corresponding sales vector to form the preference vector $\mathbf{C_{\text{sales}}}(t)$, which is used in the subsequent AIF step. Once the actual market price $P$ is observed, the SRP model is updated using the true sales from both firms, closing the loop and enabling adaptive learning over time.

\paragraph{Production cost:} 
Prior preferences over production costs encode the cost incurred by the agent when producing goods. This cost is calculated as the product of the fixed unit cost $c_i$ and the (observed) number of units produced at the previous time step:
\begin{equation}
    \mathbf{C_{\text{production}}}(t-1) = c_i \cdot q_i(t-1).
\end{equation}

\paragraph{Warehouse occupancy cost:} 
Prior preferences over warehouse occupancy assign costs based on the inferred level of stock, as indicated by the stock level signal. Higher occupancy corresponds to increased maintenance and operational costs.

\paragraph{Analysis cost:} 
Prior preferences over market analysis encode the cost of performing a warehouse signal analysis. This action reduces uncertainty by providing a noiseless signal of the warehouse state but incurs a fixed cost. It ensures that the agent balances the benefit of improved inference against the cost of acquiring it.

\subsubsection{State Prior [D]}
\label{subsec:GM_D}

Perfect perception of the initial state is assumed. Each agent therefore knows exactly its initial storage level and starting production context. In our simulations, these initial conditions are fixed: the warehouse is empty and the production context is set to \textit{acceptable production}.

\subsubsection{The Generative Process}
\label{subsec:GP_Cour}

The GP used in the numerical experiments follows a common structure across all test cases, including both the Cournot duopoly and its three-player extension. The maximum price parameter is set to $a = 30$, the demand slope to $b = 1$, the warehouse capacity to $10$ units per firm, and the maximum production per time step to $6$ units per firm. Variations between test cases affect only parameters such as the number of customers and cost structure, which are otherwise independent of the general market dynamics. This shared setup enables a consistent assessment of the proposed AIF–based methodology across diverse multi-agent configurations.

Each simulation spans 25 discrete time steps and is designed to test the robustness and adaptability of the framework under controlled evolving market conditions. The experiments begin under ideal initial conditions, where firms have accurate knowledge of the underlying market parameters. The number of customers then decreases twice during the experiment: after the 5-th and 10-th time steps, mimicking a contraction in market demand. At the 15-th time step, the market undergoes a pronounced \textit{bandwagon effect} \cite{BandwagonReview}, characterized by a simultaneous increase in both customer demand and market price. This effect reflects a collective behavioral shift commonly observed in real-world markets, where consumers’ purchasing decisions become mutually reinforcing. Customers are equally distributed between firms but are programmed to prefer buying from a competitor if possible rather than not purchasing at all. 

\subsubsection{The Active Inference Loop}
\label{subsec:AIF_loop}

An algorithmic description of the AIF loop is provided in Algorithm~\ref{alg:agent_loop}, offering a concise and transparent overview of how the proposed methods integrate within the AIF framework once the agents and the GP have been defined. The loop refers to the multi-agent system as a whole. Each agent performs inference independently within its own GM, while interactions between agents occur only through the GP, which determines the market outcomes.

\begin{algorithm}[ht]
\caption{AIF loop}
\label{alg:agent_loop}

\hspace*{\algorithmicindent}\textbf{Input:} Initialized AIF agents;\\
\hspace*{1.5em}$\textit{last\_sales}$: vector of agents' last sales\\
\hspace*{1.5em}$\textit{last\_production}$: vector of agents' last production\\
\hspace*{1.5em}$\textit{warehouse\_signal}$: signals from all warehouses\\
\hspace*{1.5em}$\textit{analysis\_signal}$: signals from performed analyses\\
\hspace*{1.5em}$\hat{a}$: vector of assumed market prices

\begin{algorithmic}[1]

\Procedure{Initialize Loop}{}

\State Set initial observations

\State Initialize SML parameters $\boldsymbol{\xi}_0$ on the null instance $(\textit{null\_sales}=[0,\dots,0],\hat a)$

\State Initialize prior estimate $\hat{a}_{\text{old}} \leftarrow \hat{a}$

\EndProcedure

\vspace{4pt}

\Procedure{Iterative Update}{}

\While{$t<T$}

\State Infer posterior beliefs over hidden states and production context

\State Update $\mathbf{C}_{\text{sales}}(t)$: for each agent $i$ and for all feasible production action $p_i$

\Statex \hspace{4em}$\widehat{P}(p_i) = f_{\text{SRP}} (p_i,\textit{last\_sales}_{\setminus i};\boldsymbol{\xi}_t)$~\eqref{eq:C_hat_estimate}

\State Infer policies and sample next actions for each agent

\State Update the GP with sampled actions

\State Extract new observations from the updated GP

\State Update the SML parameters:
\[
\boldsymbol{\xi}_{t+1} = \Psi(\boldsymbol{\xi}_t,last\_sales, \hat{a}_{\text{old}}) ~\eqref{eq:xi_update}
\]

\State Compute $\hat{a}_{\text{new}} = f_{\text{SRP}}(\textit{null\_sales}=[0,\dots,0];\boldsymbol{\xi}_{t+1})$

\If{$\frac{|\hat{a}_{\text{new}} - \hat{a}_{\text{old}}|}{\hat{a}_{\text{old}}} > 0.1$}

\State Recompute BR strategy

\State Update transition matrix $B$

\EndIf

\State $\hat{a}_{\text{old}} \leftarrow \hat{a}_{\text{new}}$

\EndWhile

\EndProcedure

\end{algorithmic}
\end{algorithm}
\subsection{Numerical Simulations}
\label{sec:Simulations}

This section presents the simulation results for the proposed AIF framework in Cournot market scenarios. The analysis is structured into three parts: first, the \textit{duopoly} setting; second, the \textit{three-firm} extension; and finally, a comparative discussion of the two cases. In both settings, we first examine the behavior of properly specified agents, each equipped with a well-defined GM, to assess their ability to handle the proposed market formulation and converge toward stable and near-optimal strategies. We then introduce an agent characterized by increased precision in the observational channel related to sales. Although similar high-precision configurations have been qualitatively associated in the literature with certain neurocognitive traits~\cite{morphogenesis}, such analogies are beyond the scope of this work. Instead, the focus is on understanding how variations in perceptual precision influence both individual behavior and collective dynamics within the presented framework.

\subsubsection{The Duopoly Scenario}
\label{subsec:2pl}

Table~\ref{tab:duopoly_params} reports the true market parameters and their corresponding estimates used by each firm in the duopoly scenario, together with the associated BR strategies. The estimated parameters $\hat{a} = a$ and $\hat{b} = b$ define each agent’s internal model of the market and are used to compute its optimal production strategy. Since $\hat{a} = a$ and $\hat{b} = b$ coincide with the true values, the computed BR strategies correspond to the Nash equilibrium of the classical one-shot Cournot game.

\begin{table}
\centering
\begin{tabular}{cccccc}
\bottomrule
\textbf{Firm} & $\hat{a}_i$ & $\hat{b}_i$ & $c_i$ (unit cost) & BR Strategy ($t < 15$) & BR Strategy ($t \geq 15$) \\
\toprule
Firm 1 & 30 & 1 & 16 & 5 & 10 \\
Firm 2 & 30 & 1 & 17 & 4 & 9 \\
\bottomrule
\end{tabular}
\caption{Internal parameters and best-response strategies for firm 1 and firm 2.}
\label{tab:duopoly_params}
\end{table}

The evolution of the number of customers in the GP is defined as:
\begin{equation}
\text{Number of customers}(t) =
\begin{cases}
10, & \text{if } t < 6, \\
6, & \text{if } 6 \leq t < 11, \\
4, & \text{if } 11 \leq t < 15, \\
15, & \text{if } t \geq 15.
\end{cases}
\end{equation}

Since customers are equally distributed across firms, both firms can sell their BR production quantities during the first six time steps. The same holds for $t\geq15$, when the theoretical BR quantity increases due to a higher market maximum price $a(t)$, defined as:
\begin{equation}\label{eq:higher-price1}
a(t) =
\begin{cases}
30, & \text{if } t < 15, \\
45, & \text{if } t \geq 15.
\end{cases}
\end{equation}

\subsubsection*{Results for Reference Generative Models}
\begin{figure}[h!]
    \centering
    \includegraphics[width=1\linewidth]{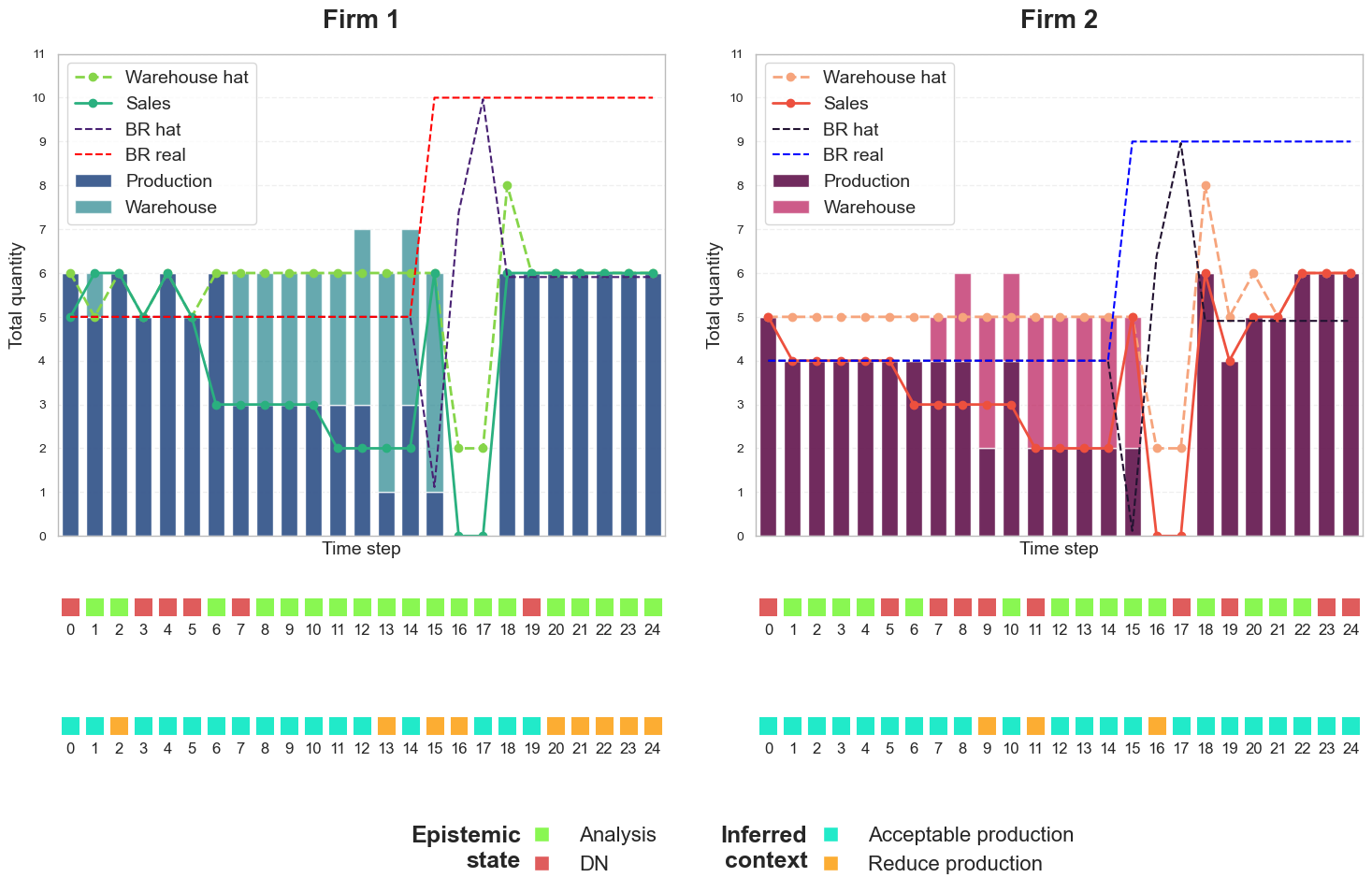}
    \caption{Behavior of firms 1 and 2 over time. Bars show the number of units produced at each time step (darker bars) and the total stock in the warehouse (lighter bars, stacked on top of production). The dashed green and orange lines represent the inferred warehouse levels for firms 1 and 2, respectively. The ``Sales'' line indicates the quantity sold at each time step; the portion of the bar above this line corresponds to unsold items carried into the warehouse. Epistemic states (\textit{DN} or \textit{Analysis}) and inferred production contexts (\textit{acceptable production} or \textit{reduce production}) are reported below. An ``Analysis'' state at time $t$ indicates that the agent performed signal analysis at time $t-1$.}
    \label{fig:Exp_Cour}
\end{figure}

Figure~\ref{fig:Exp_Cour} illustrates the outcomes of the two-firm Cournot experiment with well-designed agents. In the initial time steps, agents behave as expected. Firm~1 oscillates between producing 5 (the BR in a single-period Cournot game) and 6 units. This deviation reflects the agent’s attempt to exploit residual customer demand, as it perceives the possibility of selling one additional unit despite recognizing that producing less would yield higher profit given firm~2’s actions. Firm~2, after an initial 5 units production, quickly stabilizes around its theoretical BR of 4 units. This equilibrium persists until demand begins to decrease. As the number of customers declines, both firms start accumulating inventory. Initially, this stockpiling is not problematic, as relatively low operational costs and the potential for future sales justify maintaining higher inventory levels. However, when warehouse levels becomes excessive, the agents correctly identify a contextual shift (from \textit{acceptable production} to \textit{reduce production}) and respond by reducing output, thereby restoring stock levels within acceptable limits. In other words, when operational costs remain tolerable over a long-term policy horizon, agents sustain production; otherwise, they strategically reduce it. Qualitatively, agents aim to remain able to sell their BR quantity.

\begin{figure}[h!]
    \centering
    \includegraphics[width=0.9\linewidth]{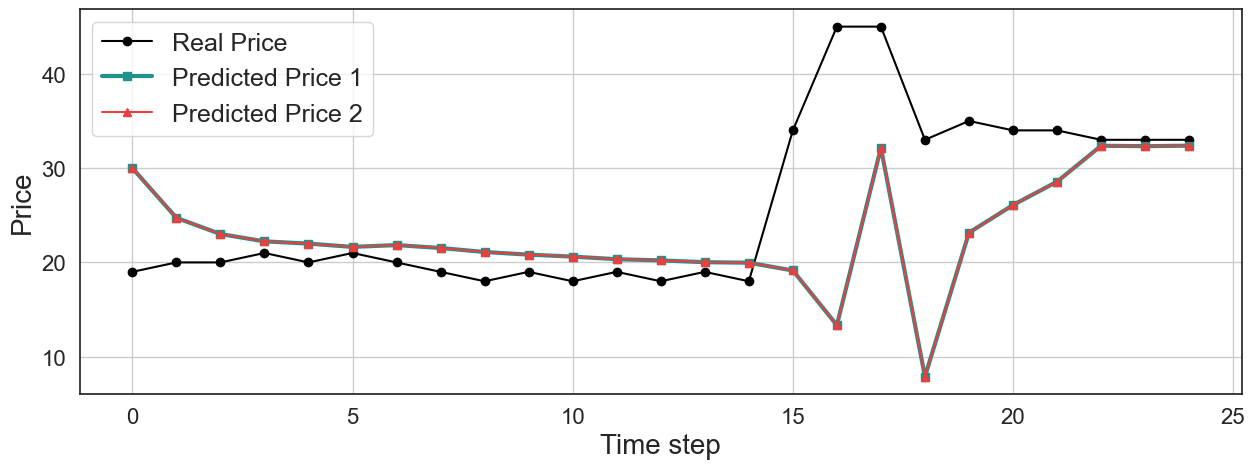}
    \caption{Real and predicted price over time. The black line represents the actual market price, while the cyan and red lines denote the predicted prices from firm~1 and firm~2, respectively. Prediction lines overlap, as both agents share the same model parameterization and learn from the same environmental information.}
    \label{fig:SML_Cour}
\end{figure}

At time step $t = 15$, as shown in Figure~\ref{fig:SML_Cour}, the market experiences a sudden price increase, accompanied by a surge in demand. This abrupt change is successfully detected by both agents, which rapidly adapt to the new conditions, leading to updated BR levels ($BR_1 = 10$, $BR_2 = 9$). Although production is constrained by each firm's operational limits, both agents increase their output to the maximum feasible level in response to the more favorable market environment. Notably, even though firm~1 infers this situation as a potential overproduction (since producing at the maximum rate generally triggers a subsequent \textit{reduce production} context), it still chooses to proceed with this strategy in light of the advantageous market conditions.

\subsubsection*{Results for Augmented Precision}

\begin{figure}
    \centering
    
    \begin{subfigure}{0.32\textwidth}
        \centering
        \includegraphics[width=\textwidth]{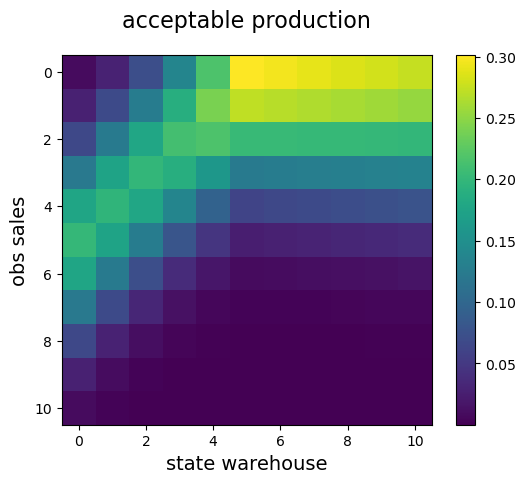}
        \caption{}
    \end{subfigure}
    \hfill
    \begin{subfigure}{0.32\textwidth}
        \centering
        \includegraphics[width=\textwidth]{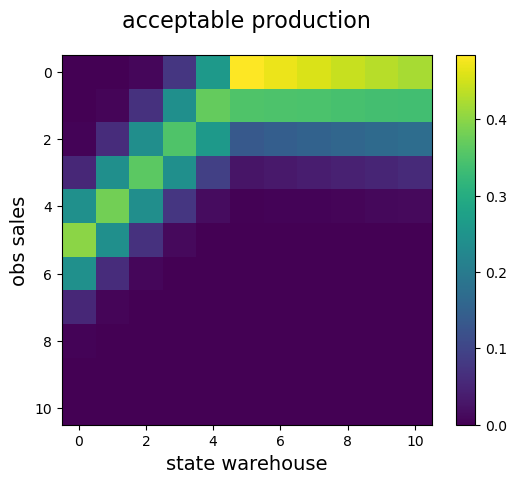}
        \caption{}
    \end{subfigure}
    \hfill
    \begin{subfigure}{0.32\textwidth}
        \centering
        \includegraphics[width=\textwidth]{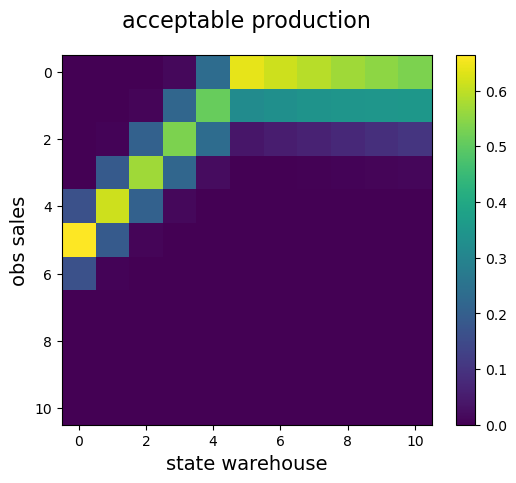}
        \caption{}
    \end{subfigure}
    \caption{Comparison of sales likelihood matrices when $\sigma$ is slightly and strongly reduced: (a) $\sigma = 2.0$; (b) slightly reduced $\sigma = 1.5$; (c) strongly reduced $\sigma = 0.6$.}
    \label{fig:sales_reduction}
\end{figure}

We now alter the framework analyzed in the previous section by reducing the variance for the sales observational channel of firm~2 from $\sigma = 2.0$ to $\sigma = 1.5$ (see  Figure~\ref{fig:sales_reduction} (a), (b)). This modification reduces the flexibility of inferring latent states given an observation. In other words, lowering the distributional variability in the likelihood matrix narrows the range of states that the agent considers plausible after observing a certain phenomenon.

\begin{figure}[t!]
    \centering
    \includegraphics[width=1\linewidth]{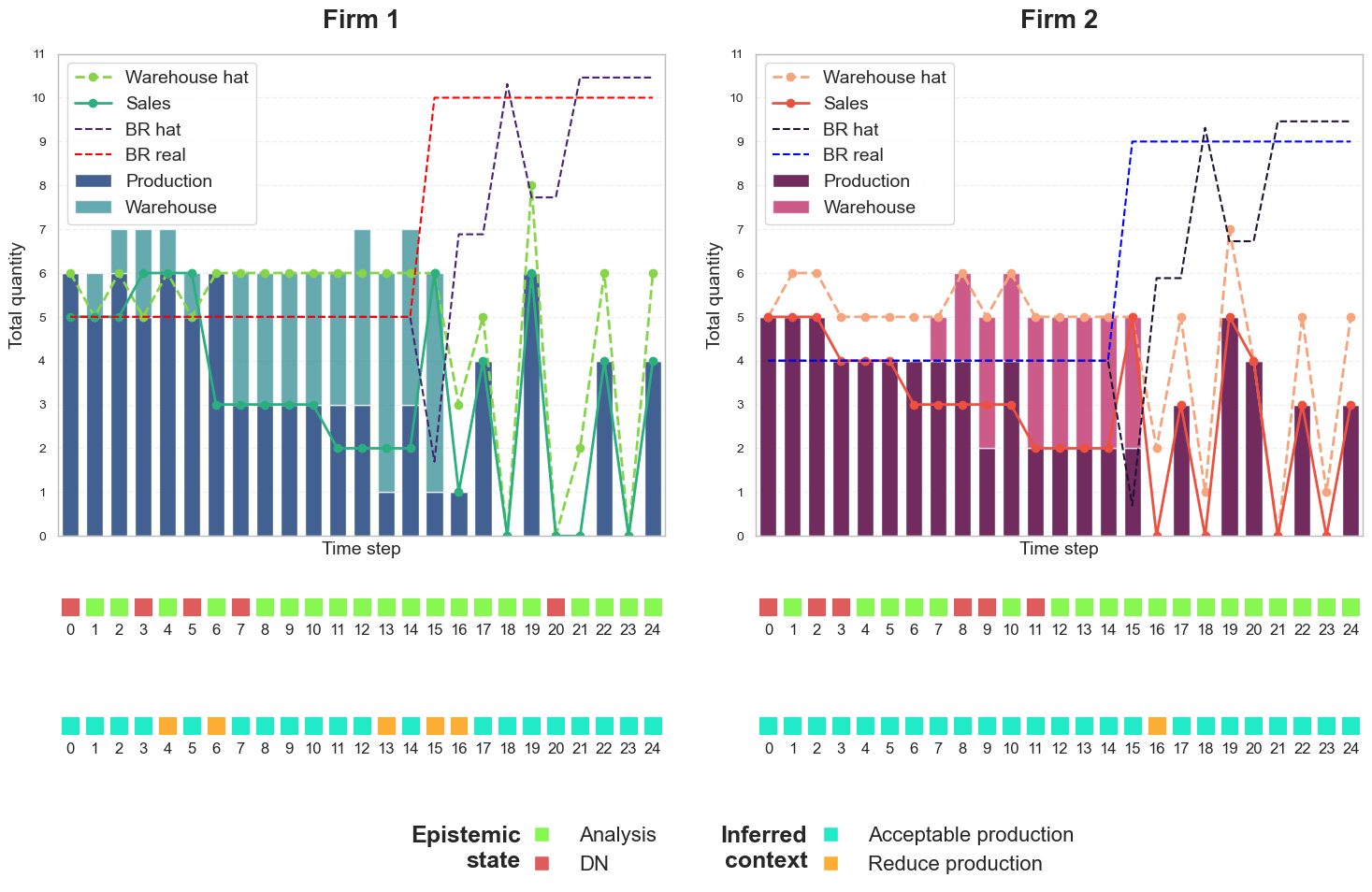}
    \caption{Cournot duopoly simulation with firm~2 exhibiting higher precision (reduced variance) in the sales observational channel.}
    \label{fig:2plautno}
\end{figure}

The dynamics of the resulting interaction is shown in Figure~\ref{fig:2plautno}. Although firm~2 (red) can adapt to initial mild variations, it fails to react effectively to the final external change. In parallel, firm~1 (blue) is ``confused’’ by the anomalous behavior of firm~2, leading to a chaotic policy selection process, even though firm~1 remains correctly parameterized. This result is attributed to the emerging complexity of the GP for firm~1, i.e., the environment becomes too unpredictable to act effectively.

To further investigate this altered behavior, we consider a simplified GP with a more stable customer demand dynamics, characterized by a single demand reduction:
\begin{equation}
\text{Number of customers}(t) =
\begin{cases}
10, & \text{if } t < 11, \\
6, & \text{if } 11 \leq t < 15, \\
15, & \text{if } t \geq 15.
\end{cases}
\end{equation}
At the same time, the variance of firm~2 is further decreased to $\sigma = 0.6$ (see Figure~\ref{fig:sales_reduction} (c)).

\begin{figure}[t!]
    \centering
    \includegraphics[width=1\linewidth]{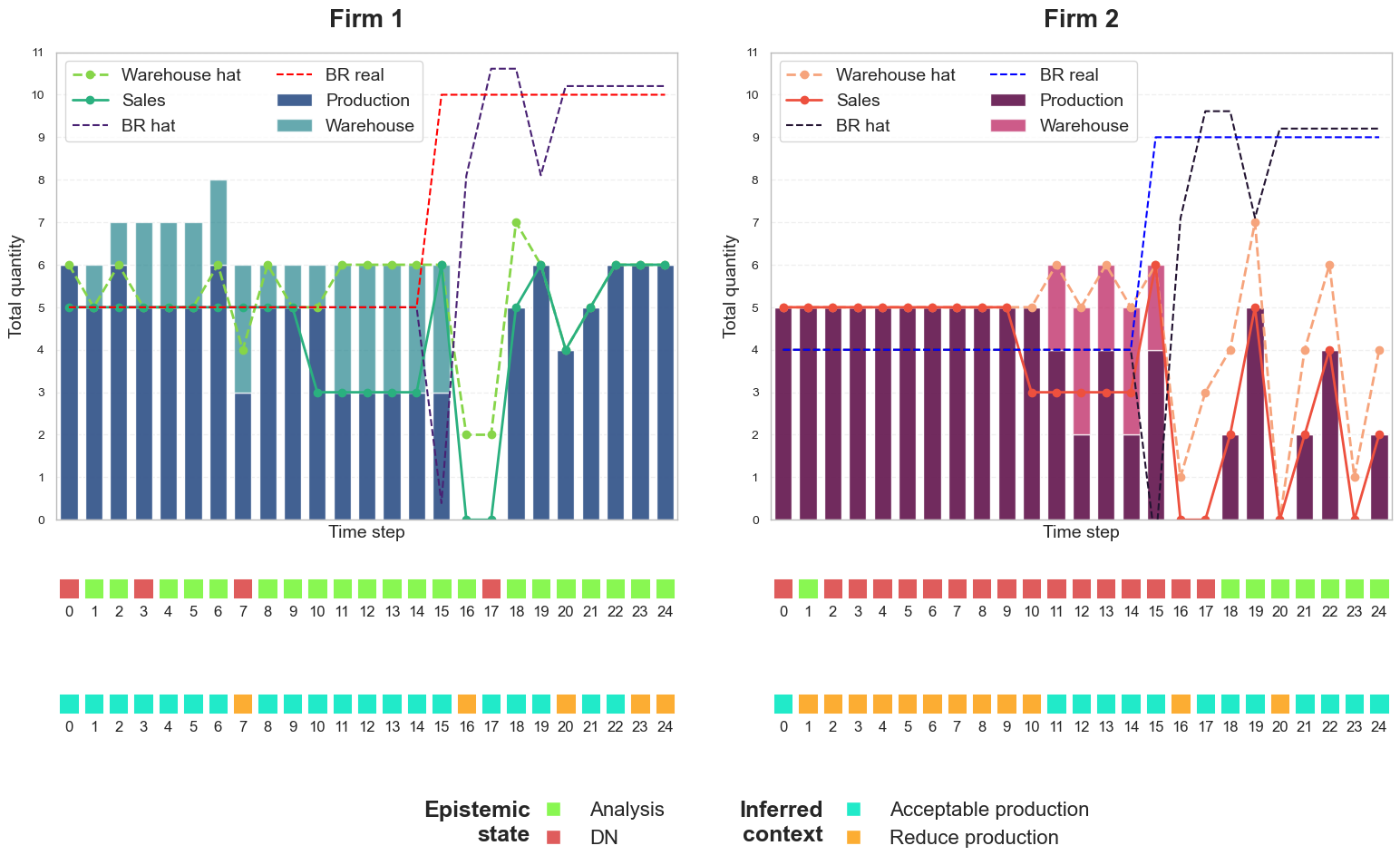}
    \caption{Cournot duopoly simulation with firm~2 exhibiting higher precision (reduced variance) in the sales observational channel and a simplified generative process.}
    \label{fig:2plautok}
\end{figure}

As shown in Figure~\ref{fig:2plautok}, both agents initially adopt sub-optimal strategies, but they remain stable. It is only when the \textit{bandwagon effect} emerges that firm~2 (red) displays a performance drop. In contrast, firm~1 (blue) retains its ability to react effectively, although it initially follows a slightly sub-optimal strategy. Compared to the reference case, firm~1 also exhibits a slight deterioration in its early warehouse management. Overall, these results indicate that firm~1 can remain self-sufficient, provided that the agent is allowed to learn over a sufficiently long time horizon.

\subsubsection{Three-Firm Extension}
Table~\ref{tab:three_firm_params} reports the parameters used in the three-firm extension of the game. Production costs are defined \textit{ad hoc} to reproduce a setting analogous to the duopoly case, enabling a consistent comparison across scenarios.

\begin{table}
\centering
\begin{tabular}{cccccc}
\toprule
\textbf{Firm} & $\hat{a}_i$ & $\hat{b}_i$ & $c_i$ (unit cost) & BR Strategy ($t < 15$) & BR Strategy ($t \geq 15$) \\
\bottomrule
Firm 1 & 30 & 1 & 6.2 & 5 & 8.75 \\
Firm 2 & 30 & 1 & 7.0 & 4 & 7.75 \\
Firm 3 & 30 & 1 & 7.8 & 3 & 6.75 \\
\bottomrule
\end{tabular}
\caption{Internal parameters and best-response strategies for the three-firm Cournot scenario.}
\label{tab:three_firm_params}
\end{table}

The ground-truth evolution of the customers number in the GP is set as follows:
\begin{equation}
\text{Number of customers}(t) =
\begin{cases}
15, & \text{if } t < 6, \\
12, & \text{if } 6 \leq t < 11, \\
9, & \text{if } 11 \leq t < 15, \\
20, & \text{if } t \geq 15.
\end{cases}
\end{equation}
Also in this case, customers are equally distributed across firms, allowing each firm to sell its BR production during the first six and the final ten time steps. The market maximum price parameter $a(t)$ follows the same dynamics  as in the duopoly case (see Eq.~\eqref{eq:higher-price1}).

\subsubsection*{Reference Generative Models}

\begin{figure}
    \centering
    \includegraphics[width=1\linewidth]{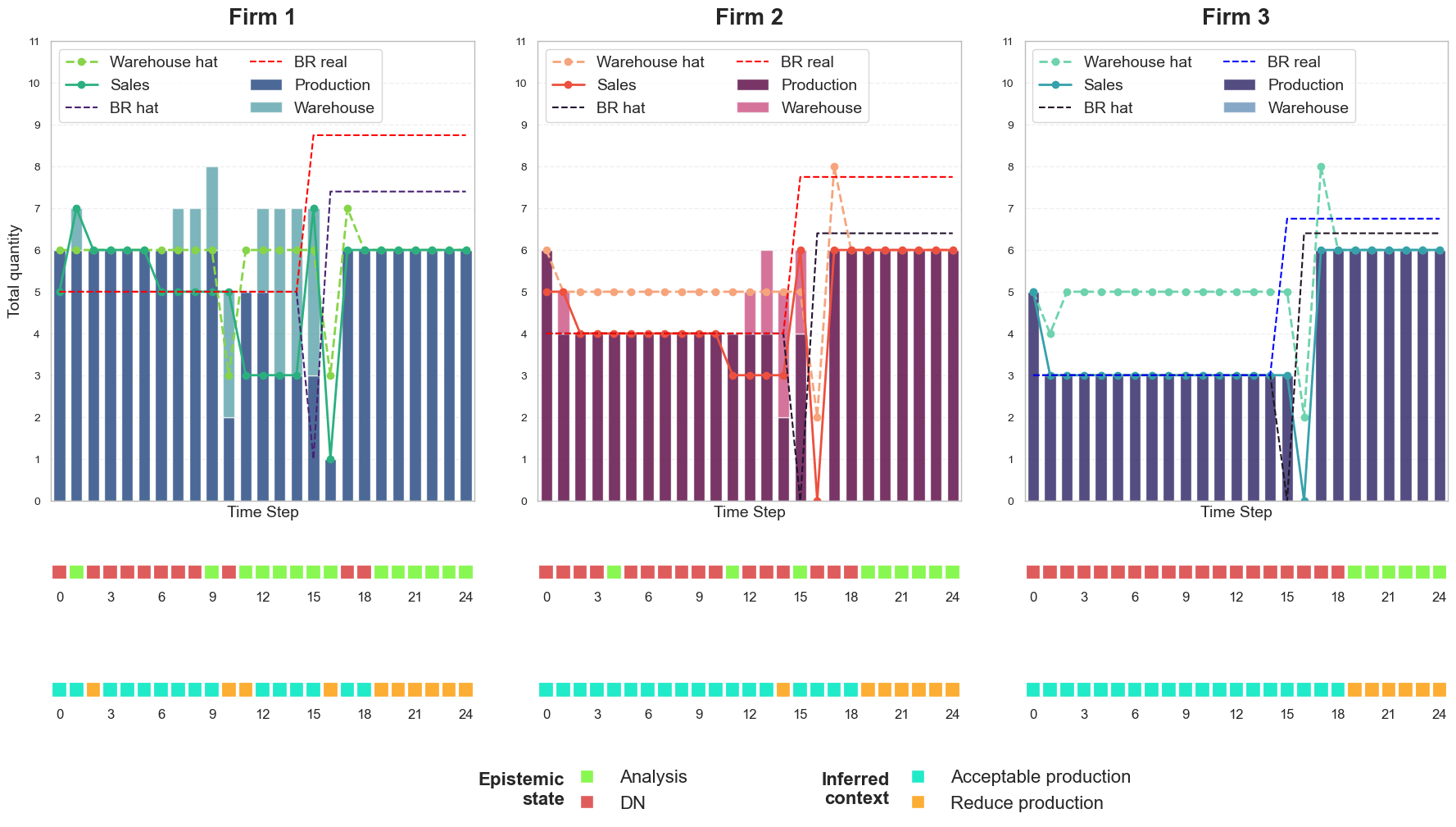}
    \caption{Cournot three-firm extension simulation with well-designed generative models.}
    \label{fig:3plok}
\end{figure}

As in the two-firm scenario, we first analyze the simulations results for equally well-designed agents. Figure~\ref{fig:3plok} shows that agents exhibit regular behavior in epistemic state transitions, context identification, and price prediction. In particular, firm~1 (the highest-producing agent by construction) requires regular analysis after the second drop in demand, when the market is perceived as too weak. Conversely, firm~3 (the lowest-producing agent) requires additional analysis only after the abrupt market change. At that point, all agents begin to interpret the situation as an overproduction context while still operating at their maximum feasible production levels, similarly to what observed in Section~\ref{subsec:2pl}. This regularity is crucial for effectively handling the higher-complexity version of the Cournot game. Moreover, as the number of agents increases, more information is naturally shared across the system. This enhanced information flow facilitates the learning process, contributing to the observed stability and coherence of agents behavior, and further highlighting the effectiveness of the AIF paradigm in multi-agent settings.

\subsubsection*{Augmenting Precision}

\begin{figure}
    \centering
    \includegraphics[width=1\linewidth]{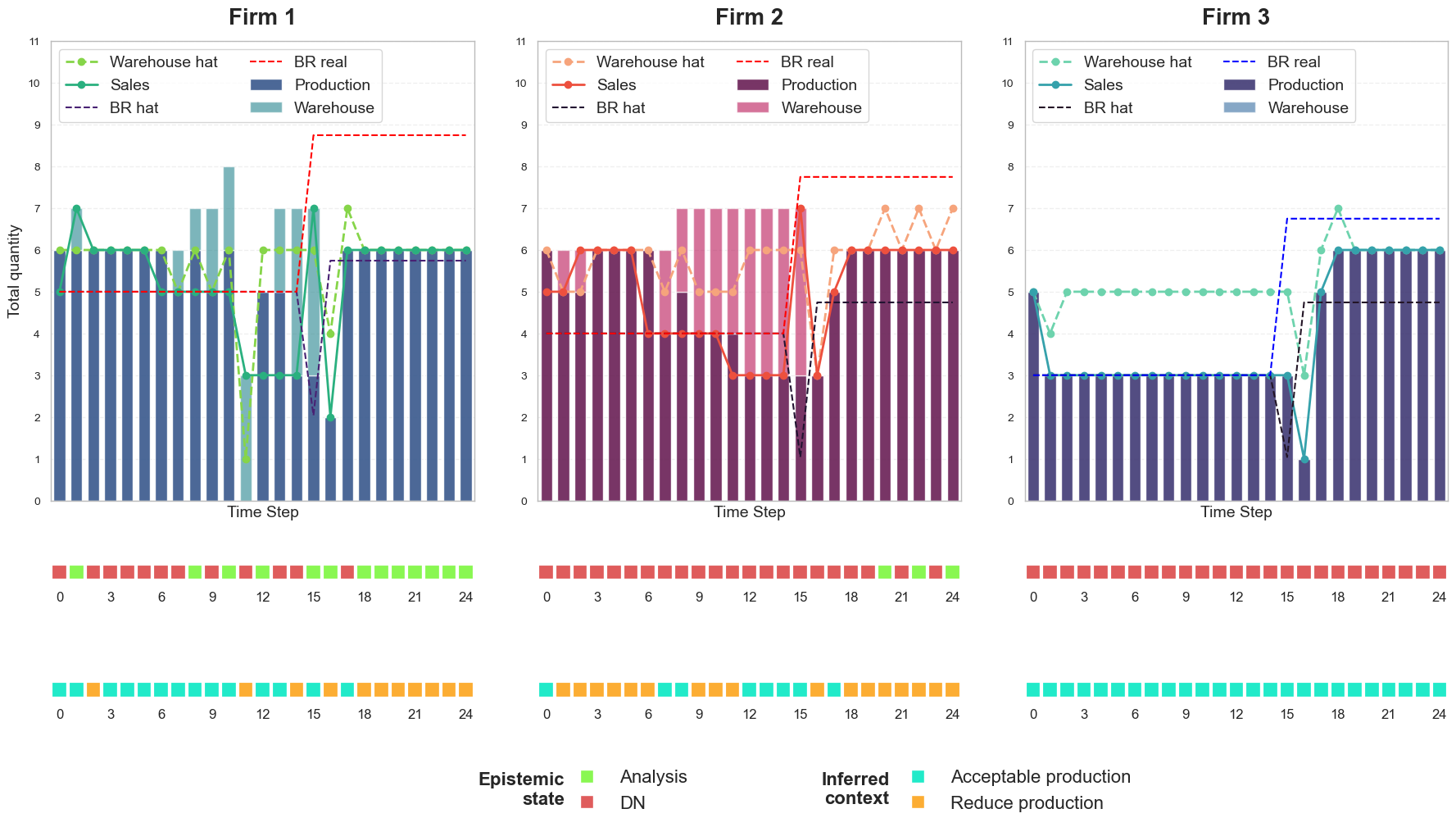}
    \caption{Cournot three-firm extension simulation with firm~2 exhibiting higher precision (reduced variance) in the sales observational channel and unchanged generative process.}
    \label{fig:3plautok}
\end{figure}

A qualitative demonstration of the previously discussed dynamics is obtained by reducing the variance of firm~2 from $\sigma = 2.0$ to $\sigma = 0.6$, without introducing any simplifications in the GP. As shown in Figure~\ref{fig:3plautok}, the high-precision agent fails to pursue its optimal BR strategy and struggles to manage inventory effectively. Nevertheless, the overall system remains stable: firms~1 and~3 preserve coherent behavior and are largely unaffected by the dynamics firm~2’s. Interestingly, and in contrast to the duopoly case (Figure~\ref{fig:2plautok}), firm~2 is still able to respond appropriately to the market shock induced by the bandwagon effect. This outcome is attributed to the stabilizing influence of the other two agents, together with the firm~2’s inherent tendency toward overproduction.

\subsection{Discussion}
\label{subsec:sim_conclusions}

The results obtained in the duopoly and three-firm scenarios highlight the effectiveness of the AIF framework in modeling multi-agent behavior and addressing well-known challenges such as the credit assignment problem. This is enabled by the explicit epistemic drive in policy selection and by the factorization of observational channels and latent states, which together provide a structured basis for context inference and adaptive behavioral modulation. When agents are equipped with well-posed GMs, they demonstrate the ability to infer hidden states, adapt to dynamic environments, and converge toward near-optimal strategies, even under non-stationary and partially observable conditions. These findings confirm that AIF provides a robust methodological and theoretical framework for capturing key aspects of decision-making in multi-agent systems. 

Furthermore, the experiments show that as the number of agents increases, the system tends to exhibit greater stability, although the influence of each individual agent diminishes. In larger systems, each agent’s ability to steer collective behavior is reduced, while inter-agent information exchange becomes a stabilizing force in the global dynamics. This emerging balance between autonomy and mutual dependency reflects the cooperative–competitive duality observed in real-world markets.

Finally, the simulations indicate that the stability of the GP plays a critical role in shaping the system dynamics. When the environment is sufficiently stable and predictable, agents are more likely to develop robust internal models and acquire the inferential capabilities needed to sustain coherent and adaptive behavior. Conversely, when the GP becomes highly variable or chaotic, individual deviations tend to amplify, allowing unstable or mis-specified agents to exert disproportionate influence on the collective dynamics.

\section{Conclusions}
\label{sec:Conclusions}

In this work, we have examined the theoretical foundations of active inference (AIF) and the behavioral properties of agents operating under this paradigm, focusing on their interaction within multi-agent systems for digital twins (DTs) applications. Framed within partially observable Markov decision processes, we have built on previous applications of AIF in collective systems \cite{knowingOnesPlace,friston2013life,maisto2025flock,Pesci,Bottoni,MT_AIF,elife,proietti2025active} and extended the paradigm to DT settings, where multiple agents interact through decentralized generative models within a shared environment.

To improve the practical applicability of AIF-based DTs, we have introduced two methodological contributions. First, contextual inference to enable agents detecting shifts in environmental conditions and adapting their behavior through context-sensitive policy selection. Second, streaming machine learning to augment generative models with an online learning component, allowing agents to handle complex or evolving environmental relationships without requiring increasingly detailed internal representations. 

Numerical experiments based on an extended Cournot competition model have illustrated how the proposed elements translate into emergent multi-agent behavior within a DT representation of a socio-economic system. Agents are able to predict future market prices without performing high-dimensional inference over both agents’ production and inventory states, providing an effective trade-off between representational simplicity and behavioral adaptability. Even in a competitive setting, agents indirectly exchange information through their shared environment, leading to increasingly stable collective dynamics as the number of agents grows.

\vspace{4pt} 
\noindent{\bf Data Accessibility:} This work builds upon a fork of the \texttt{pymdp} library \cite{pymdp}, available at \url{https://github.com/FrancescoMaria28/pymdp}, which includes the following modifications: 
(i) the introduction of the \texttt{shared\_control\_groups} option to explicitly specify which hidden states are affected by the same actions;
(ii) the introduction of the \texttt{scale\_state} option to weight the \texttt{state\_info\_gain} term. All scripts and code required to reproduce the results reported in this manuscript are available at 
\url{https://github.com/FrancescoMaria28/Active_Inference_pymdp}.

\vspace{4pt} 
\noindent{\bf Acknowledgements:} FMM and AM acknowledge the FIS Starting grant “Reduced Order Modeling and Deep Learning for the real-time approximation of PDEs (DREAM)”, Grant Agreement FIS00003154, funded by the Italian Science Fund (FIS) - Ministero dell’Università e della Ricerca. MT, AC and AM acknowledge the ERC Advanced grant IMMENSE (Grant Agreement 101140720), funded by the European Union. DM, FD, and GP acknowledge financial support from the ERC Consolidator grant ThinkAhead (Grant Agreement 820213), funded by the European Union. AM is member of the Gruppo Nazionale Calcolo Scientifico-Istituto Nazionale di Alta Matematica (GNCS-INdAM) and also acknowledges the project “Dipartimento di Eccellenza” 2023-2027 funded by MUR.

\bibliographystyle{elsarticle-num}
\biboptions{sort&compress}

\appendix

\section{Likelihood derivation for multi-state modalities}
\label{app:signal_likelihood_derivation}

This appendix details the derivation of a likelihood matrix that accounts for the two components $p(\text{warehouse signal} \mid \text{warehouse})$ and  $p(\text{warehouse signal} \mid \text{context})$. Let $w$ denote the warehouse state, $e$ the epistemic state, $c$ the context state, and $o$ the observation signal representing the level of warehouse occupancy. We assume that the observation likelihood is independent of the epistemic state, such that:
\begin{equation}
p(o \mid w, e') = p(o \mid w), \quad \forall e' \in e,
\end{equation}
and that $p(o \mid w)$ is well-defined. The goal is to compute a context-sensitive observation likelihood $p(o \mid w, c, e)$.

The quantity $p(c \mid o)$ is specified a priori and represents a mapping from observations to context. In this work:
\begin{equation}
\begin{array}{c}
p(\text{``acceptable prod.''} \mid \text{``perfect''}) \simeq 1, \quad  
p(\text{``acceptable prod.''} \mid \text{``in control''}) = 0.8, \\
p(\text{``acceptable prod.''} \mid \text{``loading''}) = 0.2, \quad 
p(\text{``acceptable prod.''} \mid \text{``out of control''}) \simeq 0, \\
1-p(\text{``acceptable prod.''} \mid o)=p(\text{``reduce prod.''} \mid o) .
\end{array} 
\end{equation}
This prior knowledge is incorporated into the observation model using Bayes' rule, as follows:
\begin{equation}
p(o \mid w, c, e) \propto p(c \mid o) \, p(o \mid w, e),
\end{equation}
which penalizes observations that are inconsistent with the given context via the term $p(c \mid o)$. Finally, normalization over all possible observations yields:
\begin{equation}
p(o \mid w, c, e) = \frac{p(c \mid o) \, p(o \mid w, e)}{\sum_{o'} p(c \mid o') \, p(o' \mid w, e)}.
\end{equation}

\section{Streaming Random Patches}
\label{sec:SRP}

Streaming random patches (SRP) is an ensemble learning algorithm designed for online learning on evolving data streams. It extends traditional batch ensemble methods by combining random sampling of instances and feature subspaces, while incorporating adaptive mechanisms to handle concept drift. The method is inspired by the random subspaces method \cite{709601}, online bagging \cite{Bagging}, and the random patches algorithm \cite{UnderstandingRF}. These ideas are integrated into the SRP ensemble through three main mechanisms: (i) \textbf{Online bagging:} each incoming instance is used to train every base learner a random number of times, sampled from a Poisson distribution. This procedure approximates bootstrap sampling in classical bagging; (ii) \textbf{Random subspaces:} each base learner operates on a randomly selected subset of features, reducing correlation between learners and increases ensemble diversity; (iii) \textbf{Drift detection:} the ensemble detects changes in the data distribution and adapts accordingly. In this work the drift detector used is ADWIN (Adaptive Windowing), which monitors the prediction error over a sliding window of observations. When a warning is raised, a background learner is initialized; if drift is confirmed, the background learner replaces the corresponding base model.

The setting considers a data stream $\mathcal{X} = \{x_t \in \mathbb{R}^n\}_{t=1}^{\infty}$ with targets $\mathcal{Y} = \{y_t \in \mathbb{R}\}_{t=1}^{\infty}$. At each time step $t$, a new instance $x_t$ arrives, the model produces a prediction $\hat{y}_t$, and the true target $y_t$ becomes available before the next instance $x_{t+1}$, enabling immediate model updates. In general, the joint data distribution may evolve over time and is denoted by $P_t(X,Y)$. Such temporal changes are referred to as concept drift. When properly detected and handled, the data stream can be viewed as a sequence of approximately stationary segments, within which standard learning assumptions remain valid.

In this work, the base learner is the Hoeffding Adaptive Tree (HAT), an incremental decision tree designed for streaming environments. HAT uses the Hoeffding bound to determine when sufficient statistical evidence is available to perform a split, allowing the tree to grow incrementally as new data arrive. It also includes mechanisms to replace outdated branches when local concept drift is detected. The training procedure of the SRP ensemble is summarized in Algorithm~\ref{alg:SRP}. Further implementation details can be found in \cite{SRP}.

\begin{algorithm}
\caption{Training procedure for Streaming Random Patches}
\label{alg:SRP}

\textbf{Inputs:} 
number of base learners $n$, 
maximum number of features per learner $m$, 
Poisson parameter $\lambda$, 
data stream $S$.

\begin{algorithmic}[1]
\State Initialize an ensemble $L$ of $n$ base learners
\State For each learner, randomly select a subset of at most $m$ features
\State Initialize an empty set of background learners $B$

\While{a new instance $(x_t, y_t)$ arrives from the stream $S$}

    \ForAll{learners $l$ in the ensemble $L$}

        \State Compute prediction $\hat{y}_t = l(x_t)$
        \State Update the performance estimate of $l$

        \State Draw $k \sim \text{Poisson}(\lambda)$
        \State Train learner $l$ with $(x_t, y_t)$ repeated $k$ times

        \If{drift detector signals a warning}
            \State Initialize a background learner $b$ with the same feature subset
            \State Add $b$ to the set $B$
        \EndIf

        \If{drift detector confirms a drift}
            \State Replace learner $l$ with its corresponding background learner
        \EndIf

    \EndFor

    \ForAll{background learners $b \in B$}
        \State Update $b$ using $(x_t, y_t)$
    \EndFor

\EndWhile

\end{algorithmic}
\end{algorithm}

\end{document}